\documentclass[showpacs,aps,preprint]{revtex4}
\usepackage{graphicx}
\usepackage{amsmath,amssymb}
\usepackage{accents}
\makeatletter
\def\widebar{\accentset{{\cc@style\underline{\mskip10mu}}}}
\makeatother

\def\gtsim{\mathrel{\hbox{\raise0.2ex
  \hbox{$>$}\kern-0.75em\raise-0.9ex\hbox{$\sim$}}}}
\def\ltsim{\mathrel{\hbox{\raise0.2ex
  \hbox{$<$}\kern-0.75em\raise-0.9ex\hbox{$\sim$}}}}

\newcommand{\bm}[1]{\mbox{\boldmath $#1$}} 

\def\mbold#1{\mbox{\boldmath $#1$}}

\begin{document}

\title{
Rotational motion of triaxially deformed nuclei studied by
microscopic angular-momentum-projection method I: \\
Nuclear wobbling motion
}

\author{Mitsuhiro Shimada, Yudai Fujioka, Shingo Tagami and Yoshifumi R. Shimizu}
\affiliation{Department of Physics, Graduate School of Science,
Kyushu University, Fukuoka 819-0395, Japan}


\begin{abstract}

Rotation of triaxially deformed nucleus has been
an interesting subject in the study of nuclear structure.
In the present series of work,
we investigate wobbling motion and chiral rotation
by employing the microscopic framework of angular-momentum projection
from cranked triaxially deformed mean-field states.
In this first part the wobbling motion is studied in detail.
The consequences of the three dimensional cranking are investigated.
It is demonstrated that the multiple wobbling rotational bands
naturally appear as a result of fully microscopic calculation.
They have the characteristic properties,
that are expected from the macroscopic triaxial-rotor model
or the phenomenological particle-triaxial-rotor model,
although quantitative agreement with the existing data is not achieved.
It is also found that the excitation spectrum
reflects dynamics of the angular-momentum vector
in the intrinsic frame of the mean-field
(transverse vs. longitudinal wobbling).
The results obtained by using the Woods-Saxon potential and
the schematic separable interaction are mainly discussed,
while some results with the Gogny D1S interaction are also presented.

\end{abstract}


\maketitle

\section{Introduction}
\label{sec:intro}

Nuclear triaxial deformation has been a long standing issue
in the field of nuclear structure~\cite{BM75}.
It is predicted that rather a small number of nuclei are
triaxially deformed in their ground states (see e.g. Ref.~\cite{Mol06}).
Near the ground state, however, it is very difficult to confirm
that a deformed nucleus has a triaxial shape.
Sometimes the existence of the $\gamma$-band
is taken as evidence of triaxiality.
However, its excitation energy,
typically about 800\,keV -- 1\,MeV in the rare-earth region,
is about an order of magnitude larger than the typical rotational (first 2$^+$)
energy and its standard interpretation is
a low-lying shape vibration around the (nearly) axially symmetric
shape~\cite{BM75}.

The situation changes at high-spin
(see e.g. Refs.~\cite{VDS83,Fra01,Pan11} and references therein).
In the region of the sizable amount of the angular momentum
the orientation of the angular-momentum vector
relative to the intrinsic nuclear shape comes into play.
Collective rotation about all three principal axes is allowed
in the triaxially deformed case so that the angular-momentum vector
can tilt from the principal axes, which leads to multiple-band structure
called wobbling~\cite{BM75}.
Note that the spectrum is quite different from much simpler ones
in nuclei with axially symmetric deformed mean-fields.
In fact, the characteristic band structure of the wobbling motion
has been measured first in the $^{163}$Lu nucleus~\cite{Ode01}
(see e.g. Refs.~\cite{NMM16,Fra17} for recent theoretical review articles).
Another specific rotational structure expected in triaxially deformed nuclei
is the appearance of chiral doublet bands, first predicted in Ref.~\cite{FM97}.
Their analysis is reported in the second part of this study.

It must be emphasized that these interesting types of rotational motion
associated with the triaxial deformation have been predicted
by phenomenological models such as
the triaxial-rotor model~\cite{BM75}
and the particle-triaxial-rotor model~\cite{FM97}.
It is certainly desirable to confirm that such rotational motion appears
by employing a fully microscopic framework,
which is the main purpose of the present work.
If necessary for the description of high-spin states,
we rely on the cranking procedure applied to
the triaxially deformed mean-field (see e.g. Ref.~\cite{Fra01}).
In order to recover the rotational symmetry of the states, we apply
the angular-momentum-projection method (see e.g. Ref.~\cite{RS80}).
Since almost all the symmetries except for the space inversion (parity)
are broken by the triaxial mean-field state in the present investigation,
an efficient method to perform the projection calculation is necessary.
We employ the method developed in Ref.~\cite{TS12},
which has been successfully applied
to the study of the nuclear tetrahedral deformation~\cite{TSD13,TSD15},
the $\gamma$-vibration~\cite{TS16},
and the ground-state rotational bands~\cite{STS15,STS16} in rare earth nuclei.

It is worth mentioning that the angular-momentum-projection technique
has been utilized with great success by the so-called projected shell model
(PSM)~\cite{HS95} (see also recent review articles~\cite{Sun16,SBD16}).
Although both our approach and the PSM rely on the angular-momentum-projection
technique, the basic philosophy is different:
In order to improve the results for high-spin states,
the multi-quasiparticle configurations are successively included
in the PSM basis with the goal of the shell-model configuration-mixing
will converge.  We optimize the mean-field states as much as possible
by the cranking procedure, which has been known to be a powerful method
for the description of high-spin states~\cite{Fra01}.
We believe that our approach provides a good alternative of the PSM,
since the cranked mean-field efficiently incorporates
the important multi-quasiparticle configurations.

The paper is organized as follows.  We briefly present our approach
in Sec.~\ref{sec:formulation}, where it is explained
how to construct the mean-field and to choose of the Hamiltonian.
In Sec.~\ref{sec:wobbling} the wobbling band in the $^{163}$Lu nucleus
is investigated.
The difference between recently proposed ``transverse'' and ``longitudinal''
wobbling~\cite{FD14} are studied in detail by our fully microscopic approach.
We should admit that we are not able to obtain good agreement with
existing experimental data in the present work.
However, we believe that it is totally non-trivial to show
that the wobbling motion naturally appears
as a result of the fully microscopic calculation.
Sec.~\ref{sec:summary} is devoted to summary of the present study.
Few preliminary results were already published in Ref.~\cite{TSF14}.

\section{Basic Formulation}
\label{sec:formulation}

The purpose of the present work is to study, with the microscopic
angular-momentum-projection method,
how the characteristic rotational features of
the triaxially deformed nuclei appear and what kind of properties they have.
For such a purpose it is preferable to be able to change
the mean-field parameters, like the deformation parameters and the pairing gaps,
arbitrarily.  Therefore, we employ a model composed of
the phenomenological Woods-Saxon potential and
a schematic separable interaction,
which has been utilized in Refs.~\cite{TS12,TSD13}.
The eigenstates of rotational band are obtained by
angular-momentum projection from the mean-field state $|\Phi \rangle$,
\begin{equation}
 |\Psi_{M\alpha}^{I}\rangle =
 \sum_{K} g_{K,\alpha}^{I}\,
 \hat P_{MK}^I|\Phi \rangle,
\label{eq:wfProj}
\end{equation}
where the operator $\hat P_{MK}^I$ is the angular-momentum projector.
The amplitudes $g^I_{K,\alpha}$ are determined by
the Hill-Wheeler equation (see e.g. Ref~\cite{RS80}),
\begin{equation}
 \sum_{K'}{\cal H}^I_{K,K'}\ g^I_{K',\alpha} =
 E^I_\alpha\,
 \sum_{K'}{\cal N}^I_{K,K'}\ g^I_{K',\alpha},
\label{eq:HW}
\end{equation}
with the definition of the Hamiltonian and norm kernels,
\begin{equation}
 \left\{ \begin{array}{c}
 {\cal H}^I_{K,K'} \\ {\cal N}^I_{K,K'} \end{array}
 \right\} = \langle \Phi |
 \left\{ \begin{array}{c} \hat H \\ 1 \end{array}
 \right\} \hat{P}_{KK'}^I | \Phi \rangle.
\label{eq:kernels}
\end{equation}
For some purpose
the properly normalized amplitudes $f^I_{K,\alpha}$ are needed
instead of the amplitudes $g^I_{K,\alpha}$,
which are defined~\cite{RS80} by
\begin{equation}
f^I_{K,\alpha}=\sum_{K'}
\bigl(\sqrt{{\cal N}^I}\,\bigr)_{K,K'}\, g^I_{K',\alpha},
\label{eq:normfNocm}
\end{equation}
where the quantity $\sqrt{{\cal N}^I}$ denotes
the square-root matrix of the norm kernel,
from which the zero-norm states are properly eliminated.

The Hamiltonian $\hat H$ in the present work is given by
\begin{equation}
 \hat H = \hat h_{\rm sph} -\frac{1}{2}\,\chi\sum_{\lambda=2,3,4}
 :\hat F^\dagger_{\lambda} \cdot \hat F^{}_{\lambda}:
 -\sum_{\tau={\rm n,p}}\sum_{\lambda=0,2} g^\tau_\lambda\
  \hat G^{\tau\dagger}_{\lambda} \cdot\hat G^\tau_{\lambda},
\label{eq:hamH}
\end{equation}
where the index $\tau$ distinguishes the neutron and proton contributions.
The spherical mean-field is composed of the kinetic energy and Woods-Saxon
potential terms,
$\hat h_{\rm sph}=\sum_{\rm \tau=n,p}\bigl( t_\tau+V_{\tau}\bigr)$,
and the particle-hole interaction is isoscalar,
$\hat F_{\lambda\mu}=\sum_{\rm \tau=n,p} \hat F_{\lambda\mu}^\tau$,
with $\hat F_{\lambda\mu}^\tau=
\sum_{ij}\langle i| F^{\tau}_{\lambda\mu} |j \rangle c^\dagger_i c^{}_j$,
while the particle-particle (or pairing channel) interaction
is given for neutrons and protons ($\tau=$n,p) separately
with $\hat G^{\tau\dagger}_{\lambda\mu}\equiv\frac{1}{2}\sum_{ij}
\langle i| G^{\tau}_{\lambda\mu} |j \rangle c^\dagger_i c^\dagger_{\tilde j}\ $
(${\tilde j}$ is the time-reversed conjugate state of $j$).
In the previous works~\cite{TS12,TSD13} we have used the different form factors
for the particle-hole and the pairing channel interactions,
while, in the present work,
we make use of a common form factor for both of them, i.e.,
\begin{equation}
 F^\tau_{\lambda\mu}(\bm{r})=
 G^\tau_{\lambda\mu}(\bm{r}) =
 R_{0\tau}\, \frac{dV_\tau^{\rm C}}{dr}\,
 Y_{\lambda\mu}(\theta,\phi),
\label{eq:FGop}
\end{equation}
where $V_\tau^{\rm C}(r)$ is the central part of
the spherical Woods-Saxon potential and $R_{0\tau}$ is its radius parameter.
We believe that this choice is more consistent,
although the final results of the angular-momentum-projection calculation
do not very much differ from the previous work,
if the force strengths are suitably chosen.
Note that the smooth cut-off of the pairing model space
for the operator $\hat G$ should be done in exactly the same way
as in the previous work (see Refs.~\cite{TS12,TSD13} for details).

The product-type mean-field state
with the pairing correlations, $|\Phi \rangle$ in Eq.~(\ref{eq:wfProj}),
is generated by the following mean-field Hamiltonian,
\begin{equation}
 \hat h_{\rm mf}=\hat h_{\rm def}
 -\sum_{\tau={\rm n,p}} p^\tau_0\,
 (\hat G^{\tau\dagger}_{00}+\hat G^{\tau}_{00})
 -\sum_{\tau={\rm n,p}} \lambda_\tau \hat N_\tau
 -\mbold{\omega}_{\rm rot}\cdot\mbold{\hat J}.
\label{eq:hammf}
\end{equation}
The first term
$\hat h_{\rm def}=\sum_{\tau} \bigl(t_\tau+V_{\tau}^{\rm (def)}\bigr)$,
is the deformed Woods-Saxon single-particle Hamiltonian
with the usual radius parameterization,
\begin{equation}
 R(\theta,\varphi)=
 R_0 \,c_v(\{\alpha\})
 \bigg[ 1+\sum_{\lambda\mu}\alpha^*_{\lambda\mu}
 Y_{\lambda\mu}^{}(\theta,\varphi) \bigg],
\label{eq:surf}
\end{equation}
which describes the deformed nuclear surface at the half depth of the potential
and the quantity $c_v(\{\alpha\})$ guarantees volume conservation.
It is noted that the mean-field Hamiltonian is not fully selfconsistent
with the two-body Hamiltonian in Eq.~(\ref{eq:hamH}).
In this work, we employ $\lambda=2$ and 4 deformations with
the parameters $(\beta_2,\beta_4,\gamma)$,
where the so-called Lund convention~\cite{BR85} is used
for the sign of triaxiality parameter $\gamma$, which means,
for example,
$\langle x^2 \rangle < \langle y^2 \rangle < \langle z^2 \rangle$
for $0^\circ < \gamma < 60^\circ$.
Here $\langle x^2 \rangle$ etc. are abbreviated notations of
${\displaystyle \Bigl\langle \sum_{a=1}^A \bigl(x^2\bigr)_a\Bigr\rangle}$ etc.,
which will be also used in the following discussions.
The parameter $p^\tau_0$ in the second term in Eq.~(\ref{eq:hammf})
fixes the strength of the monopole ($L=0$) pairing potential.
The third term takes care the number conservation
on average; i.e. the chemical potential $\lambda_\tau$ is determined
such that the particle number condition,
$\langle \hat N_\tau \rangle= N_\tau$ is satisfied.
Since the form factor of the operator $\hat G^\dagger_{00}$ is
not the simple (usual) monopole-pairing operator,
$\hat G^\dagger_{00} \ne \hat P^\dagger \equiv
\frac{1}{2}\sum_i c^\dagger_i c^\dagger_{\bar i}
=\sum_{i>0} c^\dagger_i c^\dagger_{\bar i}$,
the parameter $p^\tau_0$ is not the usual pairing-gap, $\Delta_\tau$,
which corresponds to the even-odd mass difference.
Instead we utilize the average pairing-gap,
\begin{equation}
 \Delta_\tau \equiv p^\tau_0 \, \langle\hat G^\tau_{00}\rangle
 /\langle \hat P^\tau \rangle,
\label{eq:avgap}
\end{equation}
which is always uniquely related to the parameter $p^\tau_0$.
The last term in Eq.~(\ref{eq:hammf}) is
the tilted-axis cranking term~\cite{Fra93}
with three rotational frequencies,
$\mbold{\omega}_{\rm rot}=(\omega_x,\omega_y,\omega_z)$.
Since the cranking procedure is performed for any (generally tilted)
rotation axis, we can restrict the triaxial deformation parameter
to the range, $0^\circ \le \gamma \le 60^\circ$.

Once the projected wave function~(\ref{eq:wfProj})
is obtained, it is straightforward to calculate
the electromagnetic transition probabilities~\cite{RS80}.
We use no effective charge for the calculation of the $B(E2)$ values
because a large model space ($N_{\rm osc}^{\rm max}=12$) is employed
without any kind of ``core''.
The effective spin $g$-factor of $0.7\times g_{s,{\rm free}}$ is adopted
for both neutrons and protons for the calculation of the $B(M1)$ values.
In this way there is no ambiguity for the calculation of
these reduced transition probabilities.

We employ the parameter set of the Woods-Saxon potential
proposed by R.~Wyss~\cite{WyssPriv},
the values of which are listed in Ref.~\cite{SS09}.
In this reference~\cite{SS09} the wobbling motion was studied
based on the same Woods-Saxon mean-field but
with a different microscopic framework~\cite{Mar79},
the quasiparticle random-phase-approximation (QRPA),
where only the excitation energy of the one-phonon wobbling band
can be calculated microscopically.
As for the force strengths of the interaction in Eq.~(\ref{eq:hamH}),
the so-called selfconsistent value given in Ref.~\cite{BM75} is used
for the particle-hole interaction, $\chi$
(see Refs.~\cite{TS12,TSD13} for details).
For the particle-particle channels, $g^\tau_{\lambda}$,
the monopole strength $g^\tau_0$ is determined so that the selfconsistently
determined pairing parameter $p^\tau_0=g^\tau_0 \langle\hat G^\tau_{00}\rangle$
gives the proper average pairing-gap in Eq.~(\ref{eq:avgap}).
The values of the latter are set equal to
the even-odd mass difference for the ground state of even-even nucleus,
where the deformation parameters are determined
by the Woods-Saxon Strutinsky calculation of Ref.~\cite{TST10}.
For the odd-$A$ or odd-odd nuclei,
we use the average of the neighboring even-even nuclei.
The quadrupole pairing parameter $g^\tau_2$ is assumed to be
proportional to $g^\tau_0$ and the proportionality constant is
chosen to be $g^\tau_2/g^\tau_0=1.980$, which gives the correct
$2^+$ excitation energy of the ground-state rotational band
in a typical rare earth nucleus $^{164}$Er.
Thus, the Hamiltonian is not devised with the intention to describe
the wobbling bands or the chiral doublet bands.

In most of the present investigation we employ the Woods-Saxon mean-field
and the schematic interaction in Eq.~(\ref{eq:hamH}).
However, we also show some results of a more fully selfconsistent approach,
the angular-momentum projection from the mean-field obtained
by the selfconsistent Hartree-Fock-Bogoliubov (HFB) method with
the finite-range Gogny D1S interaction~\cite{D1S}.
Recently, we have developed a computer code to perform such calculations,
and it has been applied in our previous works~\cite{TSD15,STS15,STS16,TS16}.
For example, the ground-state rotational bands of
the rare-earth nuclei can be naturally reproduced~\cite{STS15,STS16}.
The selfconsistent cranking procedure is employed
just like in Eq.~(\ref{eq:hammf}), i.e.,
for $\hat{H}'=\hat{H}-\mbold{\omega}_{\rm rot}\cdot\mbold{\hat J}$
with the Gogny interaction included in $\hat{H}$.
The calculational method is exactly the same as in these references;
especially, the Slater approximation is used in order to prevent
vanishing pairing correlation for protons in the ground state,
although we can perform the calculation without this approximation.
In contrast to the Woods-Saxon potential, the mean-field is generated
from the effective interaction in the Gogny-HFB approach
and the selfconsistent potential is non-local.
Therefore it is not easy to characterize the nuclear shape
by the mean-field potential in this case:
the shape of the mean-field is specified by the density distribution.
We use the deformation parameters defined, as usual, by
\begin{equation}
 \alpha_{\lambda\mu}({\rm den})
 \equiv
 \frac{4\pi\,\langle Q_{\lambda\mu}\rangle}
 {3A\,\widebar{R}^{\lambda}} \,,
 \qquad
 \widebar{R}
 \equiv
 \sqrt{\frac{5}{3A}
 \Bigl\langle \sum_{i=1}^{A}\bigl(r^2\bigr)_i
 \Bigr\rangle},
\label{eq:defparam}
\end{equation}
where $Q_{\lambda\mu}=r^{\lambda}Y_{\lambda\mu}$
is the mass $\lambda$-pole operator.
In the same way the amount of the pairing correlation
is characterized by the average pairing gap,
\begin{equation}
\widebar{\Delta}=
 \left[- \sum_{a>b}\Delta_{ab}\kappa^*_{ab}\right]
 \left[\sum_{a>0}\kappa^*_{a\tilde{a}}\right]^{-1},\qquad
  \Delta_{ab}= \sum_{c>d}\bar{v}_{ab,cd}\,\kappa_{cd},
\label{eq:gavgap}
\end{equation}
where the quantities $\bar{v}_{ab,cd}$ and $\kappa_{ab}$
are the anti-symmetrized matrix element of the two-body interaction
and the abnormal pairing tensor, respectively~\cite{RS80}.
In order to specify the intrinsic coordinate system of the mean-field,
we impose the constraints~\cite{KO81},
$\alpha_{21}({\rm den})=\alpha_{2-1}({\rm den})=0$
and $\alpha_{22}({\rm den})=\alpha_{2-2}({\rm den})$,
and select the $xyz$ coordinate axes to satisfy
$\langle x^2 \rangle \le \langle y^2 \rangle \le \langle z^2 \rangle$
corresponding to the Lund convention of the triaxiality parameter,
$0 \le \gamma \le 60^\circ$.

In the present work, we mainly report the result for
single mean-field $|\Phi \rangle$
with finite cranking frequencies (see Eq.~(\ref{eq:hammf})).
We employ the cranking procedure for changing
the alignment and moments of inertia in our microscopic framework.
It should, however, be emphasized that the cranking frequency
is not adjustable parameter from the theoretical point of view.
It should be treated as a second generator-coordinate combined
with the angular-momentum-projection,
\begin{equation}
 |\Psi^I_{M,\alpha}\rangle =
  \int \sum_{K} g^I_{K,\alpha}(\omega_{\rm rot})\,
  \hat P^I_{MK}|\Phi_{\rm cr}(\omega_{\rm rot})\rangle \,d\omega_{\rm rot},
\label{eq:PTanz}
\end{equation}
which was originally proposed by Peierls and Thouless~\cite{PT62}.
There are three cranking frequencies for triaxial cases.
For simplicity only one dimension is shown in Eq.~(\ref{eq:PTanz}).
We have investigated this method, for the first time,
for axially deformed cases in Refs.~\cite{STS15,STS16}
with the Gogny interaction.  The result is promising.
We call it angular-momentum-projected
multi-cranked configuration-mixing method.
Practically several cranked mean-field states, e.g.,
$|\Phi_n \rangle=|\Phi(\omega_{\rm rot}^{(n)})\rangle$,
$n=1,2,\cdots,N_{\rm mf}$, are configuration-mixed,
\begin{equation}
 |\Psi_{M\alpha}^{I}\rangle =
 \sum_{n=1}^{N_{\rm mf}}\sum_{K} g_{Kn,\alpha}^{I}\,
 \hat P_{MK}^I|\Phi_n \rangle,
\label{eq:wfProjm}
\end{equation}
which is the discrete approximation of Eq.~(\ref{eq:PTanz}).
This extended formulation is as straightforward
as any generator-coordinate method (GCM),
although the numerical task becomes much ($N_{\rm mf}^2$ times) heavier
because the norm and Hamiltonian kernels should be evaluated between
these several mean-fields states.
An application of this method will be discussed in Sec.~\ref{sec:mccmgog}.

\section{Application to wobbling band}
\label{sec:wobbling}

Wobbling motion is the quantized rotational motion of the rigid rotor,
which is a characteristic collective motion for
the triaxially deformed nucleus, see, e.g., Chap.~4-5 of Ref.~\cite{BM75}.
After long-lasting experimental efforts, it has been first identified
in $^{163}$Lu~\cite{Ode01}.
Nuclei in this region, $Z \sim 70-72$ and $N \sim 92-98$, exhibit
the so-called triaxial superdeformed (TSD) bands
at high-spin~\cite{SAB90,SchP95,Hagm04}.
Therefore, we take the nucleus $^{163}$Lu as a typical example.
The Nilsson-Strutinsky calculations predict deformation
of $\epsilon_2 \sim 0.4$ and $\gamma \sim 20^\circ$~\cite{BR04,FD15},
where $\epsilon_2$ is rather constant and the parameter $\gamma$
only slightly increases as a function of angular momentum.
Converting the deformation parameters to those of the Woods-Saxon potential
in Sec.~\ref{sec:formulation}, we use mainly $\beta_2=0.42$, $\beta_4=0.02$
and $\gamma=18^\circ$ in the present work.
The values of these parameters are the same as the previous work~\cite{SS09}.
Since we do not aim at a detailed comparison with the experimental data,
the pairing gaps for neutrons and protons are chosen to be constant,
$\Delta_{\rm n}\approx\Delta_{\rm p}\approx0.5$ MeV,
because the wobbling excitation is observed at high-spin states
where the pairing correlations are considerably reduced.
The calculations are performed within the isotropic harmonic oscillator basis.
The basis states are included up to the maximum oscillator shells,
$N_{\rm osc}^{\rm max}=12$.
As for the number of mesh points for the integration with respect
to the Euler angles $(\alpha,\beta,\gamma)$
in the angular-momentum-projector~\cite{RS80},
we mainly use $N_\alpha=N_\gamma=42,\,N_\beta=80$,
however sometimes increased up to $N_\alpha=N_\gamma=68,\,N_\beta=126$
to obtain the convergent result.
For solving the Hill-Wheeler Eq.~(\ref{eq:HW}) there appears
small-norm solutions which cause difficulties, see, e.g., Ref.~\cite{RS80},
and should be discarded.  We solve the equation several times
with different norm cut-off values from $10^{-13}$ to $10^{-6}$,
i.e., the solutions that have smaller norm eigenvalues than these values
are eliminated, and adopt the reasonable result
with the smallest possible value of the cut-off values.
Here ``reasonable" means that, for example, the spectrum as a function
of the spin $I$ is smooth enough.

We would like to mention that the selfconsistently determined
triaxiality parameter of the Nilsson potential
$\gamma({\rm Nils})\sim 20^\circ$,
or of the Woods-Saxon potential
$\gamma({\rm WS})\sim 18^\circ$,
corresponds to much smaller triaxial deformation $\gamma({\rm den})$
of the density distribution for the mean-field state,
which is defined by
\begin{equation}
 \gamma({\rm den})\equiv
 \tan^{-1}\biggl[-\frac{\sqrt{2}\langle Q_{22}\rangle}{\langle Q_{20}\rangle}
  \biggr] ,
\label{eq:gammaden}
\end{equation}
see, Ref.~\cite{SSM08} for the precise definitions of these various
$\gamma$ parameters and discussion related to them.
Namely,
\begin{equation}
 \gamma({\rm Nils})\sim 20^\circ,\ \gamma({\rm WS})\sim 18^\circ
 \quad\Leftrightarrow\quad
 \gamma({\rm den}) \sim 11-12^\circ ,
\label{eq:sgammaden}
\end{equation}
for the considered large deformation of
$\epsilon_2\sim 0.4$ or $\beta_2\sim 0.42$.
We have already confirmed in Ref.~\cite{Shim16},
which will be discussed later in Sec.~\ref{sec:gogny},
that the selfconsistent HFB calculation with the Gogny D1S interaction
also gives similar triaxial deformation,
$\gamma({\rm den})\approx 11-12^\circ$.
With this relatively small value of the triaxial deformation,
the out-of-band $B(E2)$ for the excited TSD bands,
which is a characteristic quantity to identify the wobbling motion,
is considerably underestimated~\cite{SSM08,SS09} (see also Ref.~\cite{FD15}).
In Sec.~\ref{sec:largegam} we will present also
the results of angular-momentum-projection calculations
with choosing larger triaxial deformation
than this selfconsistently determined value.

\subsection{Longitudinal and transverse wobbling}
\label{sec:transwob}

Although the observed $B(E2)$ values show the expected property
for wobbling-phonon bands~\cite{Ode01,Jens02,Gorg04},
the phonon excitation energy in the Lu isotopes
decreases as a function of angular-momentum,
which is in contrast to the original prediction
of the triaxial-rotor model~\cite{BM75}.
Recently Frauendorf and D\"onau gave an interpretation~\cite{FD14}
for this decreasing behavior of the wobbling excitation energy
within the simple triaxial particle-rotor model:
The presence of the odd proton in the high-$j$ $i_{13/2}$-orbit can change
the dependence of the excitation energy on the angular-momentum.
This was already pointed out in Ref.~\cite{SMM04}
but the interpretation was not appropriate
(see also Refs.~\cite{NMM16,Fra17}).
We briefly discuss the essence below following Ref.~\cite{FD14}.

Within the simple classical approximation,
which is called ``frozen alignment'' approximation in Ref.~\cite{FD14},
the total angular-momentum vector,
($J_x,J_y,J_z$) with the aligned high-$j$ particle along the $x$ axis
of the deformed intrinsic body, satisfies the equations,
\begin{equation}
\left\{\begin{array}{l}
 J_x^2+J_y^2+J_z^2=I(I+1), \vspace*{2mm}\cr
 {\displaystyle
 \frac{(J_x-j)^2}{2{\cal J}_x} +\frac{J_y^2}{2{\cal J}_y}
 +\frac{J_z^2}{2{\cal J}_z}=E }, \end{array}\right.
\label{eq:paroteq}
\end{equation}
where the first one describes the conservation of angular-momentum
and the second is the rotor model energy.  Here the quantities
${\cal J}_x$, ${\cal J}_y$ and ${\cal J}_z$ are the moments of inertia
of the core nucleus in the intrinsic frame.
For the fixed angular-momentum $I$, energy $E$,
and the alignment $j$, the angular-momentum vector moves along the trajectory
given by the intersection of the sphere and the ellipsoid shifted by
the amount $j$ in the $x$ direction as in Eq.~(\ref{eq:paroteq}).
If the moments of inertia satisfy the condition that the alignment axis
is the axis with the largest moment of inertia,
i.e., ${\cal J}_x>$ ${\cal J}_y$, ${\cal J}_z$,
the excitation energy of the quantized wobbling motion monotonically
increases as a function of spin $I$
in the same way as for the original rotor model without the aligned particle.
In this case the angular-momentum vector precesses always
around the largest inertia axis $x$.
However, if the inertia of the alignment axis is not the largest,
for example, ${\cal J}_y>{\cal J}_x>{\cal J}_z$, it is shown that
the excitation energy first increases and then decreases
as a function of $I$~\cite{FD14}:
The one-phonon excitation energy vanishes at the critical angular-momentum,
\begin{equation}
 I_{\rm c} = j\,\frac{{\cal J}_y}{{\cal J}_y-{\cal J}_x},
\label{eq:critam}
\end{equation}
which gives a transition from the principal-axis rotation (PAC)
to the tilted-axis rotation (TAC)~\cite{MO04}.
In this case the angular-momentum vector precesses first around
the alignment axis $x$, but then its direction moves to
the largest inertia axis $y$ when the spin increases;
namely a kind of transition of the main rotation axis
from one of the principal axes to the other occurs.
An instructive argument for the mechanism of this transition
can be found in \S3.8.6 of Ref.~\cite{Fra17}.
It will be discussed that
the spin-dependence of the excitation energies reflects
this change of the direction of
the angular-momentum vector in the intrinsic frame.
Frauendorf-D\"onau called these two different cases
longitudinal and transverse wobbling, respectively, in Ref.~\cite{FD14}
to indicate the different center of the intersection trajectory
of Eq.~(\ref{eq:paroteq}) in each case.
We will show that the simple picture is indeed realized
in the fully microscopic calculation of angular-momentum projection.

\subsection{Wobbling motion in even-even core nucleus}
\label{sec:wobee}

In order to study the wobbling spectrum, we first investigate
the ``core" nucleus, i.e., the even-even neighbor $^{162}$Yb
of the odd-proton nucleus $^{163}$Lu.
A brief research of the wobbling motion in the even-even nucleus $^{164}$Er
using the angular-momentum-projection approach has been recently performed
in Ref.~\cite{TS16} in relation to the study of the $\gamma$ vibrational band.
In fact, it is expected~\cite{MJ78} that the high-spin extension
of the $\gamma$ vibrational band changes its character to wobbling motion.
In contrast to Ref.~\cite{TS16}, where the Gogny interaction has been employed,
we use in Sec.~\ref{sec:formulation}
the Woods-Saxon mean-field and the schematic interaction consistent with it.
The mean-field parameters, the deformations and the average pairing gaps,
are simply chosen to be the same as in $^{163}$Lu, which are explained
in the beginning of this section.
It is shown that the result is not very different from that in $^{164}$Er.
It should, however, be noted that this analysis is not realistic
for the nucleus $^{162}$Yb: The TSD state appears
in the yrast region by occupying the Nilsson orbits originating from
the proton $i_{13/2}$ state, which is not occupied in the Yb isotopes
in normal situations.
So the TSD states of the $^{162}$Yb in the present work
has just a meaning of the possible core state of the TSD bands of $^{163}$Lu.

\begin{figure*}[!htb]
\begin{center}
\includegraphics[width=120mm]{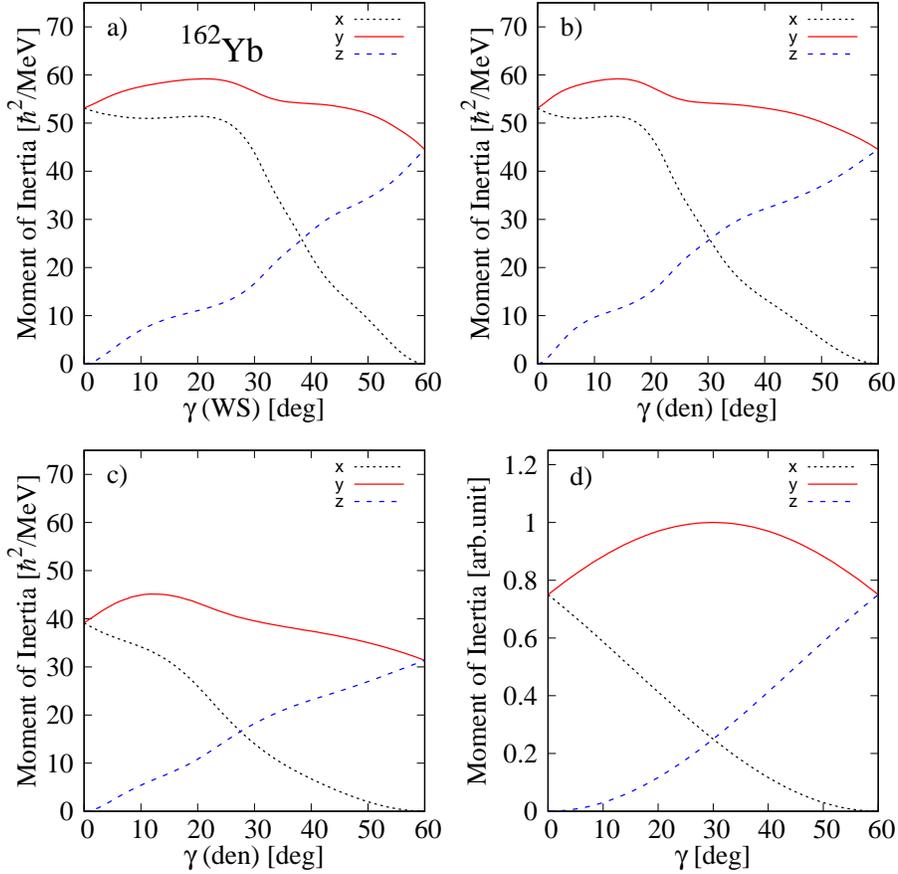}
\vspace*{-4mm}
\caption{(Color online)
Cranking moments of inertia of the three intrinsic axes, $x$, $y$, and $z$,
which are the short, medium, and long axes and denoted by dotted, solid,
and dashed lines, respectively,
as functions of the triaxiality parameter $\gamma$ for
the even-even core nucleus $^{162}$Yb of $^{163}$Lu.
The deformation parameters $\beta_2=0.42,\beta_4=0.02$ and
the pairing gaps $\Delta_{\rm n}=\Delta_{\rm p}=0.5$ MeV
are employed for the upper panels, a) and b),
and the larger pairing gaps $\Delta_{\rm n}=\Delta_{\rm p}=1.0$ MeV for c).
The $\gamma$ parameter of the Woods-Saxon potential is utilized in a) and
that of the density distribution, Eq.~(\ref{eq:gammaden}), in b) and c).
The panel~d) shows of the irrotational-flow moments of inertia
in arbitrary units.
}
\label{fig:momi}
\end{center}
\end{figure*}

As it is mentioned in the previous subsection, Sec.~\ref{sec:transwob},
the values of the moments of inertia of the core nucleus are very important
to interpret the wobbling excitation energy (see e.g. Eq.~(\ref{eq:paroteq})).
In the fully microscopic framework of the present work,
the moments of inertia are not introduced explicitly
but they should be extracted from the resultant spectrum of calculation.
However, there is no unique way to relate the calculated energy spectrum
to the three moments of inertia.
We show the cranking moments of inertia~\cite{RS80} in Fig.~\ref{fig:momi}
as functions of the $\gamma$ deformation.
They are calculated as ${\cal J}_i
={\displaystyle \mathop{\rm lim}_{\omega_i \rightarrow 0}}
\langle J_i \rangle/\omega_i$,
where $\omega_i$ is the cranking frequency about
the $i$-th axis ($i=x,y,z$) in Eq.~(\ref{eq:hammf}).
As it is noticed in Eq.~(\ref{eq:sgammaden}) the values of the triaxiality
parameter specifying the Woods-Saxon potential shape $\gamma({\rm WS})$
and that of the density distribution $\gamma({\rm den})$ are considerably
different, so that the inertias as functions of $\gamma({\rm WS})$
are shown in Fig.~\ref{fig:momi}~a),
while those as functions of $\gamma({\rm den})$ in Fig.~\ref{fig:momi}~b).
The moments of inertia calculated with larger pairing gaps,
$\Delta_{\rm n}=\Delta_{\rm p}=1.0$ MeV,
which roughly correspond to the ground-state values,
are also shown in Fig.~\ref{fig:momi}~c).
As for reference the macroscopic irrotational-flow inertias
are also included in Fig.~\ref{fig:momi}~d).
It can can be seen that the largest inertia is that of the medium
axis, i.e., the one of the $y$ axis for $0 < \gamma < 60^\circ$,
as in the case of irrotational flow.
The calculated moments of inertia
as functions of $\gamma({\rm den})$ resemble more
the irrotational-flow inertias than those as functions of $\gamma({\rm WS})$,
which is usually used to specify the triaxiality of the Woods-Saxon potential.
The $\gamma$ parameter of the irrotational moments of inertia
is naturally interpreted as that of the density distribution.
However, the relative values of three calculated moments of inertia
are considerably different from those of irrotational flow.
For example, at $\gamma=30^\circ$, the irrotational-flow ${\cal J}_y$
is four times larger than ${\cal J}_x={\cal J}_z$,
while the microscopically calculated ${\cal J}_y$
is only about two times larger.
Moreover, the symmetry with respect to $\gamma=30^\circ$ is not present
in the microscopic cranking moments of inertia.
It is known that the values of the cranking inertia are generally different
for the prolate and oblate shapes even with the same $\beta_2$ deformation.
The ``unnatural" bump-like behavior of the calculated inertia ${\cal J}_x$
at $\gamma({\rm WS})\approx 24^\circ$ ($\gamma({\rm den})\approx 16^\circ$)
is due to a sharp level-crossing of the neutron single-particle routhians
at the Fermi surface (see e.g. Fig.~1 of Ref.~\cite{SS09}).
Thus, the moments of inertia ${\cal J}_x$ and ${\cal J}_y$
at the triaxiality parameter in Eq.~(\ref{eq:sgammaden}) are comparable
and satisfy ${\cal J}_y \gtsim {\cal J}_x \gg {\cal J}_z$.
An estimated value for the critical angular-momentum in Eq.~(\ref{eq:critam})
at $\gamma({\rm WS})=18^\circ$ is rather large,
$I_{\rm c}\approx 50.1$, for $j=13/2$.
It should be noticed that these values of the inertias at the triaxiality
in Eq.~(\ref{eq:sgammaden}) are very similar to those in Ref.~\cite{FD14}.

\begin{figure*}[!htb]
\begin{center}
\includegraphics[width=155mm]{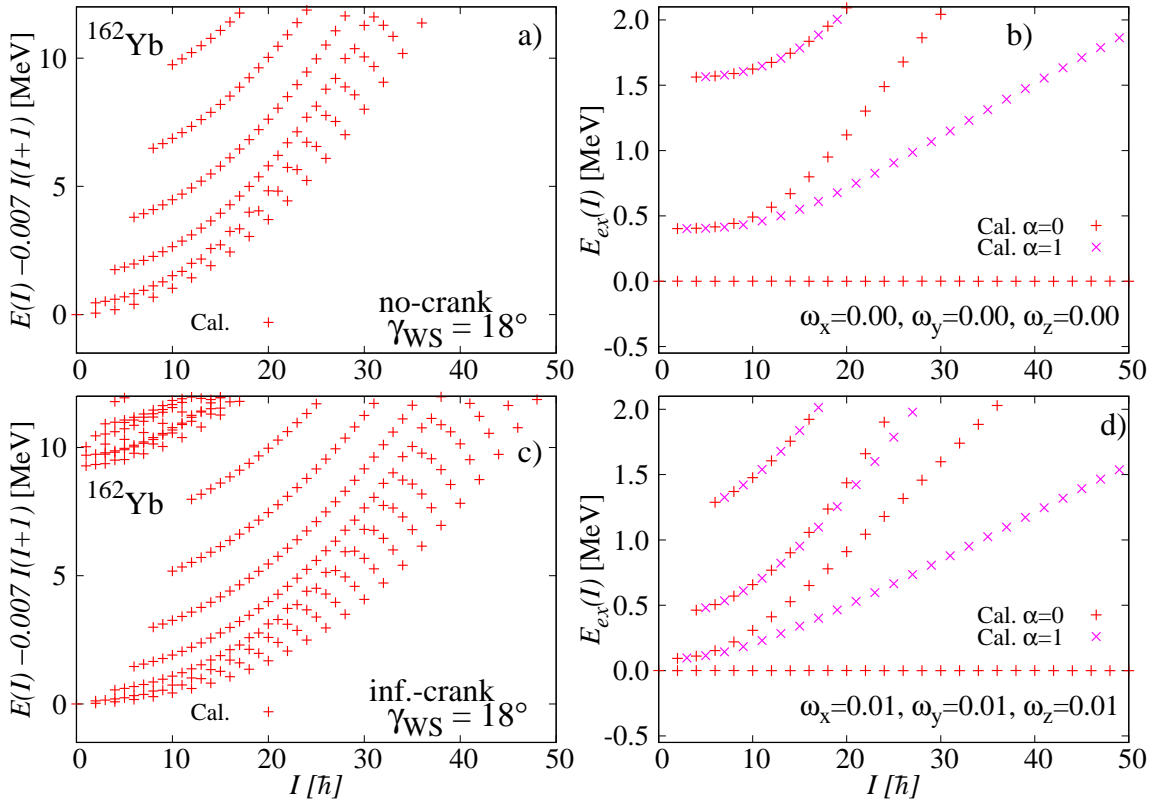}
\vspace*{-4mm}
\caption{(Color online)
The calculated spectrum (left panels) and
the relative excitation energies (right panels)
for $^{162}$Yb obtained by the angular-momentum-projection method.
The rigid-rotor reference energy $0.007\,I(I+1)$ MeV is subtracted
in the left panels.
The upper panels are the result of projection from the non-cranked mean-field,
while the lower panels from the infinitesimally cranked mean-field,
with $\omega_x=\omega_y=\omega_z=0.01$ MeV$/\hbar$.
\label{fig:wYbnoinfcr}
}
\end{center}
\end{figure*}

Now we show in Fig.~\ref{fig:wYbnoinfcr} the spectrum calculated by
angular-momentum projection from the Woods-Saxon mean-field
with the deformation parameters specified in the beginning of this section,
especially the $\gamma$ deformation parameter in Eq.~(\ref{eq:sgammaden}).
No cranking is performed for the mean-field state in the upper panels.
The lowest energy is chosen to be the energy origin ($E(I=0)=0$),
and the rigid-rotor reference energy $0.007\, I(I+1)$ MeV is subtracted
here and in the following calculations.
The value $0.007$ roughly corresponds to the one obtained by
the moment of inertia of the observed TSD1 band in $^{163}$Lu.
As it is clearly seen in the figure, the spectrum shows
the multiple-band structure characteristic for the wobbling motion.
Note that only one rotational band appears
if the mean-field is axially symmetric.
In the right panels of the figure the excitation energies of the lowest
bands are depicted as functions of the angular momentum.
The excitation energies of the excited bands
increase just as they are expected for even-even nuclei~\cite{BM75}.
The first excited band has odd spin and
the second excited band has even spin etc..
The signature of the excited band changes alternatively.
At the low-spin states the first and the second excited bands are
almost degenerate and compose the $\gamma$-band-like structure,
$2^+$, $3^+,$ $4^+$,$\cdots$, but the degeneracy is lifted after $I \gtsim 10$.
Note that we always refer the spin values $I$ in units of $\hbar$.
It should be mentioned, however, that the calculated moment of inertia
for the yrast band is rather small,
about ${\cal J}^{(1)}\approx 32$ $\hbar^2/$MeV at $I\approx 20$,
where the first moment of inertia is defined by
\begin{equation}
 {\cal J}^{(1)}(I) = \frac{(2I+1)\hbar^2}{E(I+1)-E(I-1)}.
\label{eq:J1}
\end{equation}
Compared with moment of inertia obtained for
the observed TSD1 band of $^{163}$Lu, the calculated value is less than half.
It has been pointed out that it is important to include the time-odd components
into the wave function to obtain the correct moment of inertia.
The easiest way to incorporate them is to use small frequency cranking.
We called it ``infinitesimal cranking'' in Ref.~\cite{TS16},
where it has been shown that the excitation energy
of the $\gamma$ vibration and the moments of inertia both
of the ground-state band and of the $\gamma$-band are improved.
We show the result of the projection from the infinitesimally cranked
mean-field with $\omega_x=\omega_y=\omega_z=0.01$ MeV$/\hbar$
in the lower panels in Fig.~\ref{fig:wYbnoinfcr}.
Here the small cranking frequency 10 keV is chosen to include
the time-odd contributions without the higher order effects.
The result is independent of the particular choice of this value
(see Ref.~\cite{TS16} for the proof).
Comparing the upper and lower panels of Fig.~\ref{fig:wYbnoinfcr},
the moment of inertia for the yrast band is increased
by infinitesimal cranking.  However, the value of
${\cal J}^{(1)}\approx 44$ $\hbar^2/$MeV at $I\approx 20$,
is still considerably smaller than experimental one
even if one considers the fact that the effect of alignment is present
for ${\cal J}^{(1)}$ of $^{163}$Lu.
Therefore, we are mainly concerned about the excitation energies
from the yrast band.
The effect of infinitesimal cranking is also large for the excitation energies
of the multiple wobbling bands, and the energies of the first and second
excited bands start to split at lower spin values.
It is, however, noted that the basic feature of the multiple wobbling bands
are the same; e.g., the excitation energies increase as functions of spin.

\begin{figure*}[!htb]
\begin{center}
\includegraphics[width=155mm]{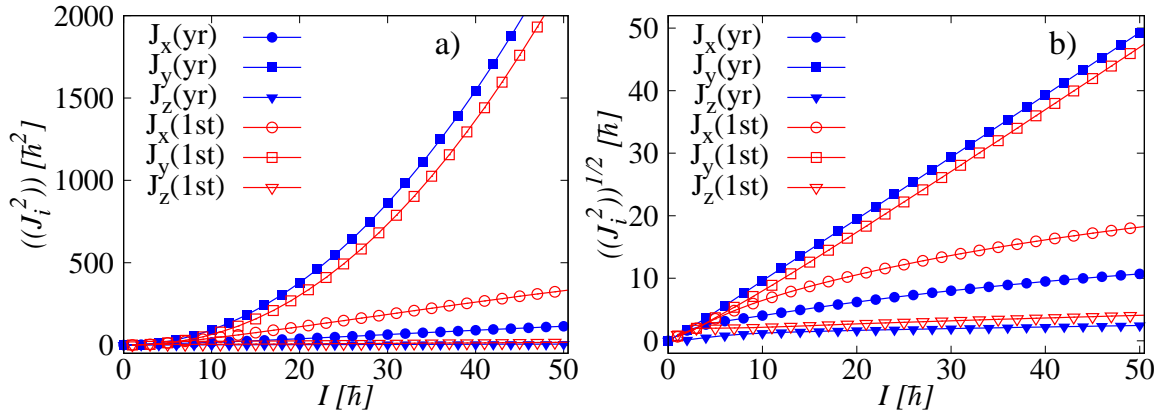}
\vspace*{-4mm}
\caption{(Color online)
The calculated behavior of the angular-momentum vector in the body-fixed frame
for the yrast and the first excited bands in $^{162}$Yb.
Here the mean-field is infinitesimally cranked
with $\omega_x=\omega_y=\omega_z=0.01$ MeV$/\hbar$
corresponding to the lower panels of Fig.~\ref{fig:wYbnoinfcr}.
The left panel shows the expectation values
of the components of squared angular-momentum operator
defined by Eq.~(\ref{eq:exJJ}),
while the right panel shows the square-root of them.
}
\label{fig:jYbinfcr}
\end{center}
\end{figure*}

In order to study the dynamical motion of the angular-momentum vector,
we consider the expectation values of the angular-momentum vector
in the body-fixed frame specified by the mean-field,
from which the projection is performed.
The expectation values of the components of
the angular-momentum vector in the intrinsic frame
are not well-defined quantities for
the angular-momentum projected wave-function in Eq.~(\ref{eq:wfProj}).
We follow the previous work~\cite{TS16} and define them
for the projected eigenstate $\alpha$ in the following way,
\begin{equation}
 (\!( J^2_i )\!)_\alpha
 \equiv \sum_{KK'} f^{I*}_{K,\alpha}\,
 \langle IK|J^2_i|IK'\rangle\,f^I_{K',\alpha},
\label{eq:exJJ}
\end{equation}
where the index $i=x,y,z$ denotes the axis
specified by the deformed intrinsic mean-field wave function $|\Phi\rangle$,
and the amplitudes $(f^I_{K,\alpha})$ are the properly normalized ones
in Eq.~(\ref{eq:normfNocm}) obtained by the projection calculation.
Needless to say, the purely algebraic quantity $\langle IK|J^2_i|IK'\rangle$,
e.g., $\langle IK|J^2_z|IK'\rangle=\delta_{KK'}K^2$,
should be calculated in the intrinsic frame with $[J_x,J_y]=-i\hbar J_z$ etc.
The microscopic geometrical information is contained
in the amplitudes $f^I_{K,\alpha}$.
A more microscopic definition by using the mean-field wave function
with the projection operator can be introduced,
which is shown to be consistent with the definition above,
see the discussion in Appendix of Ref.~\cite{TS16}.
We show the result for the infinitesimal cranking case
in Fig.~\ref{fig:jYbinfcr}, which
corresponds to the lower panels of Fig.~\ref{fig:wYbnoinfcr}.
The result for the no-cranking case is very similar and is not shown.
As it can be seen from the behavior of the three
moments of inertia in Fig.~\ref{fig:momi},
the angular-momentum vector precesses mainly
about the largest-inertia axis, i.e., the medium ($y$) axis,
and tilts slightly to the second-large-inertia axis,
i.e., the short ($x$) axis,
for both the yrast and the first excited bands.
For the first excited band the vector more tilts
to the direction of the $x$ axis,
with essentially no components along the smallest-inertia axis,
i.e., the long ($z$) axis, for both the yrast and the first excited bands.
This behavior of the expectation values of the angular-momentum vector
in the body-fixed frame are very similar to the case of $^{164}$Er
studied in the previous work~\cite{TS16}.

\begin{figure*}[!htb]
\begin{center}
\includegraphics[width=155mm]{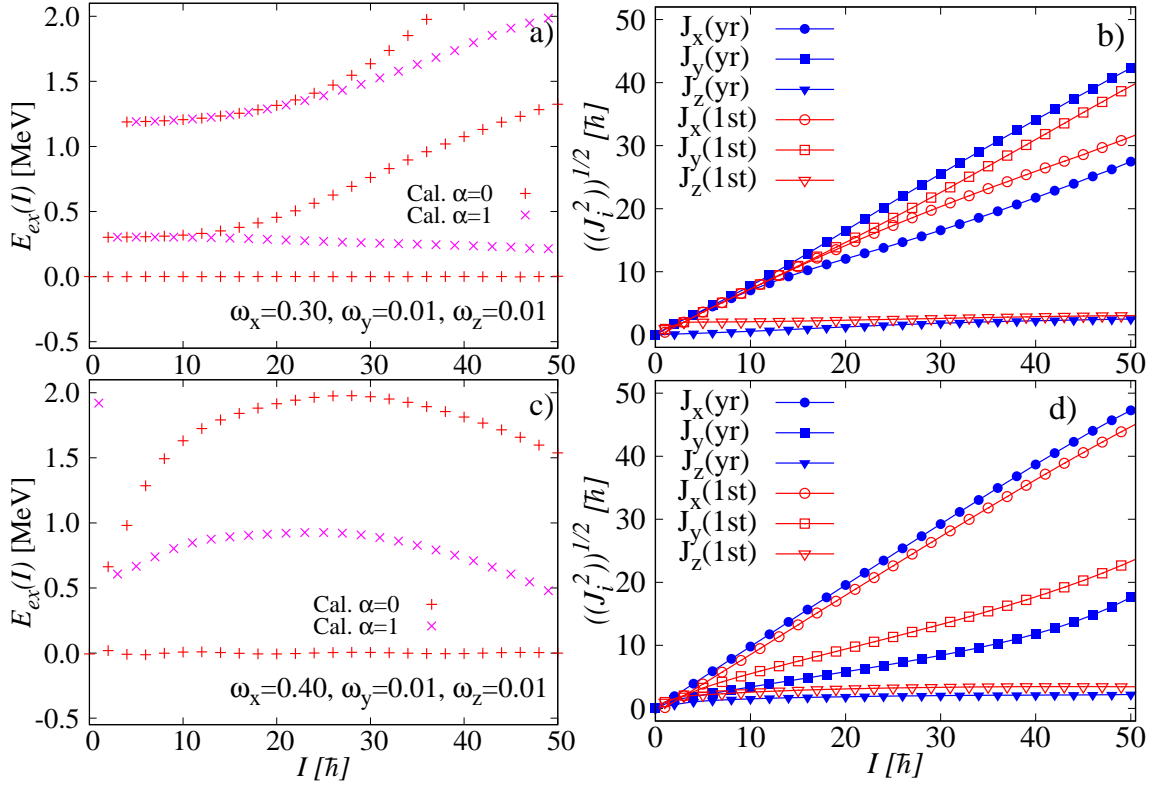}
\vspace*{-4mm}
\caption{(Color online)
The calculated excitation energy (left panels) and
the expectation values of the angular-momentum vector (right panels)
in $^{162}$Yb calculated by the projection from the cranked mean-field
with the frequencies
$\omega_x=0.30$, $\omega_y=\omega_z=0.01$ MeV$/\hbar$ (upper panels),
and those with the frequencies
$\omega_x=0.40$, $\omega_y=\omega_z=0.01$ MeV$/\hbar$ (lower panels).
}
\label{fig:ejYb0304}
\end{center}
\end{figure*}

\begin{figure}[!htb]
\begin{center}
\includegraphics[width=75mm]{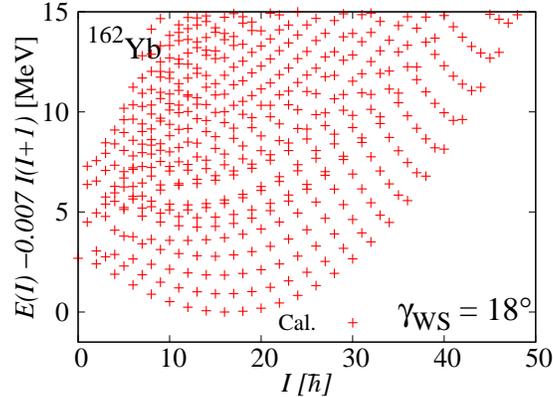}
\vspace*{-4mm}
\caption{(Color online)
The spectrum of the angular-momentum projection obtained from
the cranked mean-field with the frequencies
$\omega_x=0.40$, $\omega_y=\omega_z=0.01$ MeV$/\hbar$ in $^{162}$Yb,
corresponding to the lower panels of Fig.~\ref{fig:ejYb0304}.
}
\label{fig:sYb04}
\end{center}
\end{figure}

We study the wobbling spectrum obtained
by the angular-momentum projection from the cranked mean-field.
In the upper and lower panels of Figure~\ref{fig:ejYb0304},
we show the excitation energies and
the expectation values of angular-momentum vector for the case
with $\omega_x=0.30$ MeV$/\hbar$, $\omega_y=\omega_z=0.01$ MeV$/\hbar$ and
with $\omega_x=0.40$ MeV$/\hbar$, $\omega_y=\omega_z=0.01$ MeV$/\hbar$,
respectively.
For the case with $\omega_x=0.30$ MeV$/\hbar$, there is still no sharp alignment
in the cranked wave function, but the mean-field contains a considerable
amount of collective rotational angular-momentum of
$\langle J_x \rangle \approx 16.5\,\hbar$.
Therefore, the components of the angular-momentum vector for the $x$ and $y$
axes are comparable and then the excitation energy of the first excited band
does not increase but is almost constant as a function of spin.
Just after $\omega_x=0.30$ MeV$/\hbar$ the two $i_{13/2}$ proton
quasiparticles align their angular momenta along the $x$ axis,
(see, e.g., the quasiparticle energy diagram, Fig.~2 in Ref.~\cite{SS09}),
and the mean-field expectation value jumps to
$\langle J_x \rangle \approx 33\,\hbar$ for $\omega_x=0.40$ MeV$/\hbar$
(the collective angular-momentum also contributes when increasing
the rotational frequency from $\omega_x=0.30$ to $0.40$ MeV$/\hbar$).
Thus, the expectation value of the angular-momentum vector has
the largest component for the $x$ axis in this case
as it is seen in the lower-right panel of Fig.~\ref{fig:ejYb0304},
and the behavior of the excitation energy completely changes from
that of the lower cranking frequencies
in the upper panel of Fig.~\ref{fig:ejYb0304} or in Fig.~\ref{fig:wYbnoinfcr}.
Apparently, in this case, the excitation energy first increases and
then gradually decreases as a function of spin,
which resembles the behavior of the transverse wobbling.
However, the whole spectrum looks also very different from those in the case of
the non-cranked or of the infinitesimally cranked mean-field
in Fig.~\ref{fig:wYbnoinfcr}, as it is displayed in Fig.~\ref{fig:sYb04},
where the lowest state is not $I=0^+$ state but $I=16^+$ state,
because of the large aligned angular-momentum along the short ($x$) axis.
In this way, the effect of alignment changes
the dynamical behavior of the angular-momentum vector in the body-fixed frame,
which seems to be reflected to the wobbling-phonon excitation energy.
We will confirm this interesting relation in more detail in the following
for the odd nucleus $^{163}$Lu.

\subsection{Wobbling motion in odd nucleus $^{163}$Lu}
\label{sec:wobodd}

For the study of the wobbling motion in the odd nucleus $^{163}$Lu,
it is important to recognize which orbit the odd proton occupies.
With the deformation $(\beta_2,\beta_4)=(0.42,0.02)$ and the
triaxiality parameter in Eq.(\ref{eq:sgammaden}), the positive parity
proton orbit at the Fermi surface originates from the high-$j$
$i_{13/2}$ particle, which strongly favors to align its angular momentum
along the short axis, i.e., the $x$ axis for $0 < \gamma < 60^\circ$.
In fact, the occupation of this orbit strongly polarizes nucleus
to have sizable positive-$\gamma$ triaxial deformation
(see, e.g., Refs.~\cite{SAB90,SchP95,BR04}).
Considering that the even-even core nucleus has the moments of inertia
satisfying ${\cal J}_y \gtsim {\cal J}_x \gg {\cal J}_z$
(see Fig.~\ref{fig:momi}),
the condition for the transverse wobbling discussed
in Sec.~\ref{sec:transwob} is satisfied.

To construct the mean-field state with odd proton number,
one has to block a proton quasiparticle~\cite{RS80}.
When the cranking procedure is employed,
there is no ambiguity because the cranked quasiparticle energies
are non-degenerate, and
the lowest proton quasiparticle state is blocked.
In the case of no cranking, however, there is the Kramers two-fold degeneracy.
The mean-field state is ambiguous because any linear combinations
of the two degenerate quasiparticle states can be taken to construct it.
It should be stressed that this ambiguity of the mean-field
with the odd particle number
causes no problem for angular-momentum projection;
i.e., the projected energy is unique.
This is because the two independent quasiparticle states are transformed by
the $\pi$-rotation around one of the coordinate axes into each other,
and therefore they produce exactly the same spectrum
by the angular-momentum projection (c.f., the identity,
$\hat P_{MK}^I \hat R|\Phi \rangle=\hat P_{MK}^I| \Phi \rangle$,
where $\hat R$ is a rotational operator at any Euler angle).
We have numerically confirmed this fact.

\begin{figure*}[!htb]
\begin{center}
\includegraphics[width=155mm]{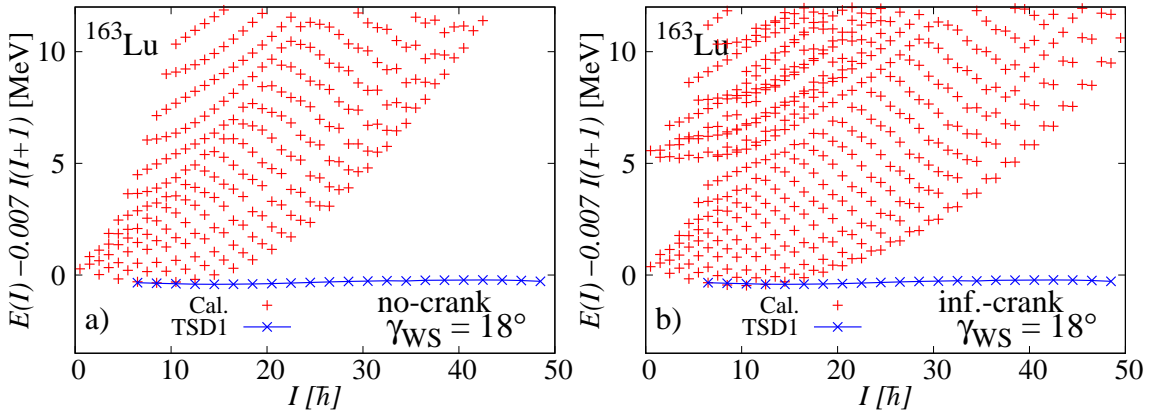}
\vspace*{-4mm}
\caption{(Color online)
Wobbling spectrum for $^{163}$Lu calculated by
the angular-momentum-projection method from
the non-cranked mean-field (left panel),
and from the infinitesimally cranked mean-field
with the frequencies
$\omega_x=\omega_y=\omega_z=0.01$ MeV$/\hbar$ (right panel).
The rigid-rotor reference energy $0.007\,I(I+1)$ MeV is subtracted
for both the calculated and experimental energies.
The energy of the experimental TSD1~\cite{Jens02} is also included
in each panel.
}
\label{fig:sLunoinfcr}
\end{center}
\end{figure*}

We first show in Fig.~\ref{fig:sLunoinfcr} the calculated spectrum
of angular-momentum projection from the non-cranked mean-field
(left panel) and from the infinitesimally cranked mean-field
with $\omega_x=\omega_y=\omega_z=0.01$ MeV$/\hbar$ (right panel),
where the experimental energy of the TSD1 band is also included.
The rigid-rotor reference energy $0.007\,I(I+1)$ MeV is subtracted
from both the calculated and experimental energies.
The lowest energy state, whose energy is chosen to be the origin (zero-energy),
has $I^\pi=9/2^+$ for no cranking (left panel),
while it has $I^\pi=13/2^+$ for infinitesimal cranking (right panel).
The experimental data for the lowest band, i.e., the TSD1 band,
is also shown, in which the lowest observed state has $I^\pi=13/2^+$.
It is confirmed that the lowest (yrast) band has the signature $\alpha=+1/2$,
the first excited band $\alpha=-1/2$, etc.;
the signatures of the excited bands change alternatively
as it is expected for the wobbling motion with a $\pi i_{13/2}$ odd nucleon.
In both cases the multiple-band structure expected
for the wobbling motion appears naturally.
However, the calculated moments of inertia of the wobbling bands are
considerably underestimated compared with those of the experimental TSD1 band:
The calculated ${\cal J}^{(1)}$ in Eq.~(\ref{eq:J1})
for the non-cranked mean-field is about 38 $\hbar^2/$MeV at $I\approx 20$,
in contrast to the experimentally measured value about 69 $\hbar^2/$MeV.
Infinitesimal cranking improves the situation,
${\cal J}^{(1)}\approx 46 \hbar^2/$MeV at $I\approx 20$, though not enough,
which is similar to the results of the even-even core nucleus $^{162}$Yb.
Some improvement for the microscopic Hamiltonian and/or some adjustment of
the interaction strengths may be necessary to reproduce
the experimental spectrum,
which is out of the scope of the present investigation.
Therefore, we mainly concentrate on
the excitation energies of the wobbling-phonon bands.

It may be worth mentioning that an extra multiple-band structure appears
at higher excitation energy $\gtsim 5$ MeV, when infinitesimally cranked
in the right panel of Fig.~\ref{fig:sLunoinfcr}.
It is interpreted as wobbling bands excited on some of higher
quasiparticle configurations that are included by infinitesimal cranking.
A similar structure at higher excitation energy $\gtsim 9$ MeV is also seen
for $^{162}$Yb in the lower-left panel in Fig.~\ref{fig:wYbnoinfcr}.
Such excited wobbling structures appear in the result of projection
from cranked mean-fields,
but how many they are and what excitation energies
they have depends on each case.

\begin{figure*}[!htb]
\begin{center}
\includegraphics[width=155mm]{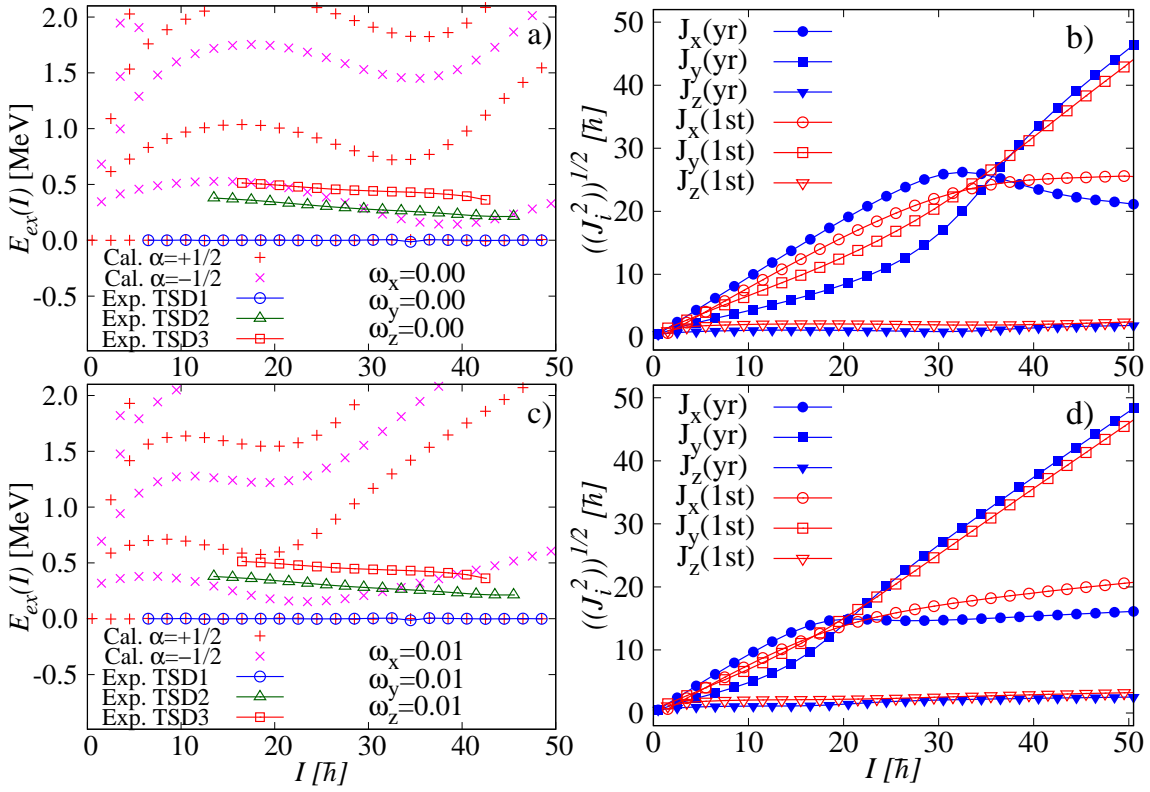}
\vspace*{-4mm}
\caption{(Color online)
The calculated excitation energy (left panels) and
the expectation values of the angular-momentum vector (right panels)
in $^{163}$Lu calculated by the projection
from the non-cranked mean-field (upper panels),
and those from the infinitesimally cranked mean-field with
$\omega_x=\omega_y=\omega_z=0.01$ MeV$/\hbar$ (lower panels).
The experimental excitation energies of TSD1, TSD2, and TSD3~\cite{Jens02}
are also included in the left panels.
}
\label{fig:ejLunoinfcr}
\end{center}
\end{figure*}

Figure~\ref{fig:ejLunoinfcr} displays
the calculated excitation energies and
the expectation values of the angular-momentum vector
in the intrinsic frame for the case with the non-cranked
and the infinitesimally cranked mean-fields
in the upper and lower panels, respectively.
The experimental excitation energies are also included in the left panels.
It can be seen from the left panels that
the characteristic features of the transverse wobbling
are realized in these calculations.
Namely, the excitation energy of the one-phonon wobbling band
first increase and then decrease as spin increases,
and it vanishes at the critical angular-momentum,
$I_{\rm c}\approx 36$ ($I_{\rm c}\approx 20$),
in the upper-left (lower-left) panel of Fig.~\ref{fig:ejLunoinfcr}.
The excitation energy does not exactly vanishes but
the energies of the lowest and the first excited bands
repel with each other, i.e., there is a virtual crossing.
Comparing the upper and lower panels in Fig.~\ref{fig:ejLunoinfcr},
the infinitesimal cranking reduces the excitation energies
and therefore the critical angular-momentum of
the vanishing one-phonon wobbling excitation energy becomes lower.
Moreover, the main component of the angular-momentum vector
in the intrinsic frame is along the alignment axis (short axis),
$x$ axis, at low spins,
and it changes to be along the largest-inertia axis, $y$ axis, at high spins.
As expected, the spin value,
where the main component exchanges from that of $x$ axis to that of $y$ axis,
$ (\!( J^2_x )\!)^{1/2} \approx (\!( J^2_y )\!)^{1/2}$,
almost corresponds to the critical angular-momentum.
Note that the collective rotation takes place around the short ($x$) axis
at low-spin states induced by
the alignment of $\pi i_{13/2}$ particle,
even if the core moments of inertia satisfy
${\cal J}_x \ltsim {\cal J}_y$ (see Fig.~\ref{fig:momi}):
The maximum values of $(\!( J^2_x )\!)^{1/2}$ are about 27
and 15 in the upper and lower panels of Fig.~\ref{fig:ejLunoinfcr},
respectively, which are much larger than the maximum alignment
of one $i_{13/2}$ quasiparticle, $j=13/2$.
This behavior of angular-momentum vector is consistent with
transverse wobbling in the classical model in Eq.~(\ref{eq:paroteq})
(see also the discussion in Ref.~\cite{Fra17}).

\begin{figure*}[!htb]
\begin{center}
\includegraphics[width=155mm]{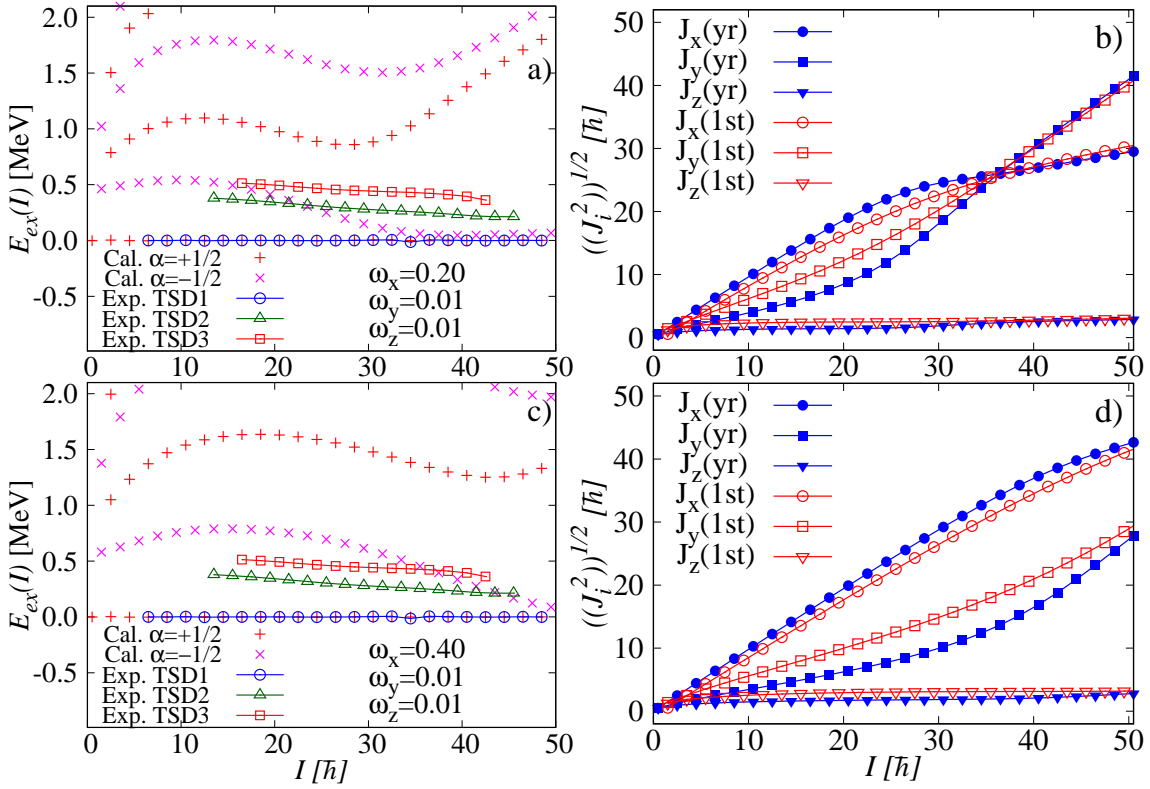}
\vspace*{-4mm}
\caption{(Color online)
The calculated excitation energy (left panels) and
the expectation values of the angular-momentum vector (right panels)
in $^{163}$Lu calculated by the projection from the cranked mean-field
with the frequencies
$\omega_x=0.20$, $\omega_y=\omega_z=0.01$ MeV$/\hbar$ (upper panels),
and those with the frequencies
$\omega_x=0.40$, $\omega_y=\omega_z=0.01$ MeV$/\hbar$ (lower panels).
}
\label{fig:ejLu0204}
\end{center}
\end{figure*}

This correspondence between the wobbling-phonon excitation energy
and the behavior of the angular-momentum vector in the intrinsic frame
seems to be rather general.   Figure~\ref{fig:ejLu0204} shows
the result of calculation using cranked mean-fields
with higher rotational frequencies,
$\omega_x=0.20$ and $0.40$ MeV$/\hbar$ in the upper and lower panels,
respectively, on top of the infinitesimal cranking.
As it is seen, the wobbling-phonon excitation energy increases
by increasing the rotational frequency around the alignment axis ($x$ axis),
and the critical angular-momentum of the vanishing one-phonon energy
becomes higher.  This is consistent with the macroscopic particle-rotor model
of the transverse wobbling referred to in Sec.~\S\ref{sec:transwob};
the higher cranking frequency $\omega_x$ effectively increases
the alignment and/or the moment of inertia about the $x$ axis,
so that the critical angular-momentum
$I_{\rm c} = j(1-{\cal J}_x/{\cal J}_y)^{-1}$ delays.
Compared with the experimental data, the one-phonon excitation energy decreases
too quickly as a function of spin.   Moreover, the two-phonon excitation
energy is almost double of the one-phonon energy, which is too large
in comparison with the data.  The experimental TSD3 excitation energy
is much smaller than the double of the TSD2 excitation energy.
Thus, our results of the projection calculation are not very successful
to reproduce the experimental spectrum.

\begin{figure*}[!htb]
\begin{center}
\includegraphics[width=155mm]{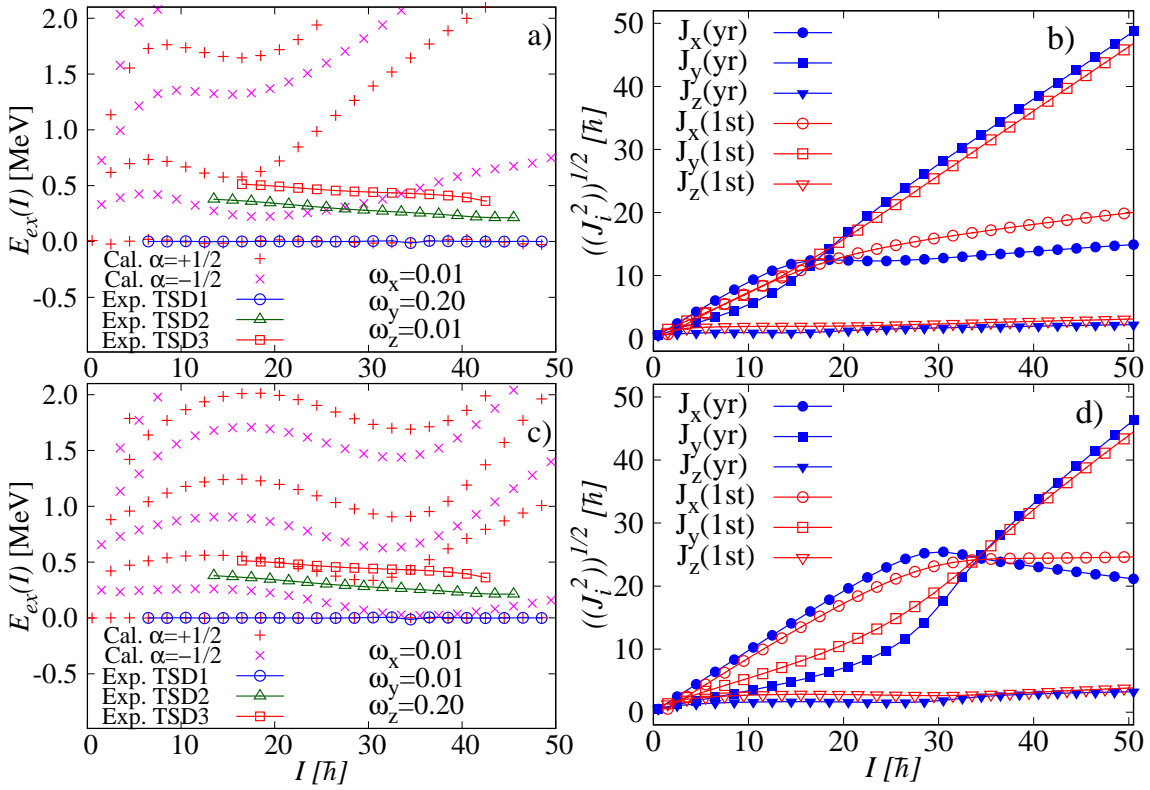}
\vspace*{-4mm}
\caption{(Color online)
The calculated excitation energy (left panels) and
the expectation values of the angular-momentum vector (right panels)
in $^{163}$Lu calculated by the projection from the cranked mean-field
with the frequencies
$\omega_x=0.01$, $\omega_y=0.20$, $\omega_z=0.01$ MeV$/\hbar$ (upper panels),
and those with the frequencies
$\omega_x=\omega_y=0.01$, $\omega_z=0.20$ MeV$/\hbar$ (lower panels).
}
\label{fig:ejLu02yz}
\end{center}
\end{figure*}

In this way it is demonstrated that the wobbling excitation energy
is sensitive to the dynamics of the angular-momentum vector
in the intrinsic frame, which is controlled by cranking of the mean-field.
The decrease of the one-phonon energy is related to the change of the main
component of the angular-momentum vector, i.e., the decrease of
$ (\!( J^2_x )\!)^{1/2} - (\!( J^2_y )\!)^{1/2}$, which changes sign
near the critical angular-momentum of the vanishing excitation energy.
We show two more examples, where the cranking with the frequency
$0.2$ MeV/$\hbar$ is performed around
the largest- and smallest-inertia axes ($y$ and $z$ axes)
in the upper and lower panels of Fig.~\ref{fig:ejLu02yz}, respectively,
on top of the infinitesimal cranking.
As it is seen, the result of the cranking around the largest-inertia axis
in the upper panels is not very different from that of
the simple infinitesimal cranking
in the lower panels of Fig.~\ref{fig:ejLunoinfcr}.
Although the cranking around $y$ axis increases $(\!( J^2_y )\!)^{1/2}$,
the frequency, $\omega_y=0.2$ MeV/$\hbar$, is not large enough
to change the alignment pattern:
The main rotation axis is still the short axis at low spins,
and only the critical angular-momentum becomes smaller,
which is also consistent with the simple model estimate,
$I_{\rm c} = j(1-{\cal J}_x/{\cal J}_y)^{-1}$.
The excitation energies of the wobbling motion look like those
of the longitudinal wobbling at spins higher than the critical one.
On the other hand, cranking around the long axis
in the lower panels of Fig.~\ref{fig:ejLu02yz} changes the result of projection
in a different way compared with the lower panels of Fig.~\ref{fig:ejLunoinfcr}.
The critical angular-momentum is increased but
the excitation energies of the wobbling-phonon become considerably smaller,
and the multiple-band structure is more clearly exhibited.
The critical angular-momentum is similar to the result of cranking around
the intermediate-inertia axis ($x$ axis)
in the upper panels of Fig.~\ref{fig:ejLu0204},
while the wobbling excitation energies are considerably smaller.
In the case of the $z$ axis cranking,
the component of angular-momentum along the largest-inertia axis
($y$ axis) is more reduced at lower spins as it can be seen by comparing
the lower panels of Figs.~\ref{fig:ejLu02yz} and
the upper panels of Figs.~\ref{fig:ejLu0204}.
It is interesting that such a difference between the expectation values
of the angular-momentum vector in the intrinsic frame is clearly
reflected in the wobbling excitation energies.

\begin{figure*}[!htb]
\begin{center}
\includegraphics[width=155mm]{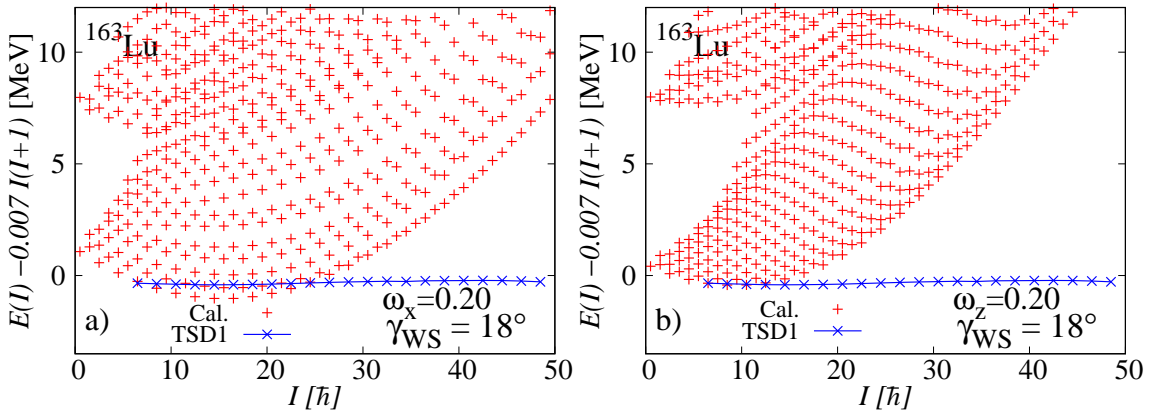}
\vspace*{-4mm}
\caption{(Color online)
Wobbling spectrum for $^{163}$Lu calculated by
the projection from the cranked mean-field
with $\omega_x=0.20$, $\omega_y=\omega_z=0.01$ MeV$/\hbar$,
corresponding to the upper panels of Fig.~\ref{fig:ejLu0204} (left panel),
and with $\omega_x=\omega_y=0.01$, $\omega_z=0.20$ MeV$/\hbar$,
corresponding to the lower panels of Fig.~\ref{fig:ejLu02yz} (right panel).
The energy of the experimental TSD1 is also included in both panels.
}
\label{fig:sLuss02}
\end{center}
\end{figure*}

It is demonstrated that the cranking of the mean-field around the $x$ axis
increases the wobbling excitation energies
while the cranking around the $z$ axis reduces them.
However, the cranking procedure also change the slope of the rotational
spectrum, i.e., the moment of inertia of the rotational band.
We compare the spectrum calculated with cranking around the $x$ axis
and the $z$ axis in Fig.~\ref{fig:sLuss02}.
The cranking around the $z$ axis makes
the moment of inertia of the TSD1 band considerably smaller,
${\cal J}^{(1)}\approx 38 \hbar^2/$MeV at $I\approx 20$,
as it is displayed in the right panel in comparison with
the cranking around the $x$ axis in the left panel,
in this case, ${\cal J}^{(1)}\approx 53 \hbar^2/$MeV at $I\approx 20$.
Therefore, the cranking only around the long axis ($z$ axis)
is not favorable for high-spin states.

\begin{figure*}[!htb]
\begin{center}
\includegraphics[width=155mm]{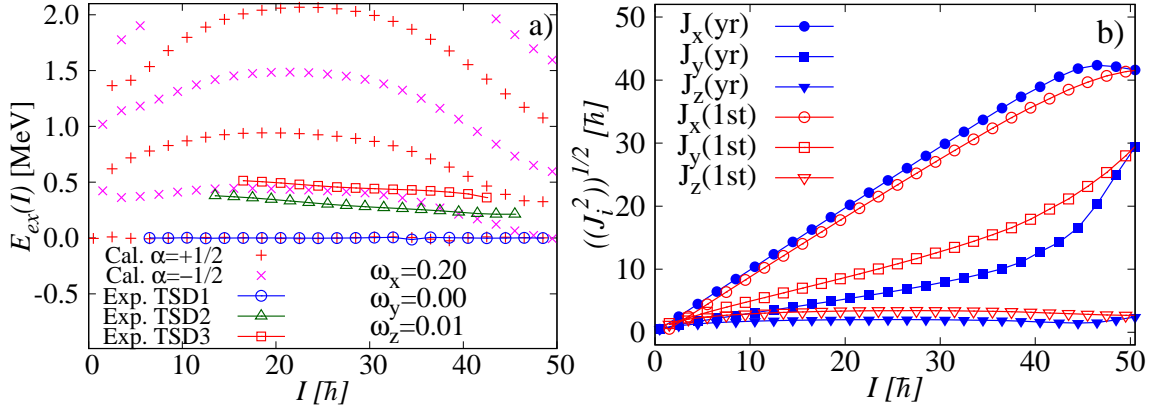}
\vspace*{-4mm}
\caption{(Color online)
The calculated excitation energy (left panel) and
the expectation values (right panel)
of the angular-momentum vector for the cranked mean-field
with $\omega_x=0.20$, $\omega_y=0.0$, $\omega_z=0.01$ MeV$/\hbar$ in $^{163}$Lu.
The experimental excitation energies of TSD1, TSD2, and TSD3
are also included in the left panel.
}
\label{fig:ejLu02std}
\end{center}
\end{figure*}

In comparison with the experimental data, all the results presented
for the wobbling excitation energies,
Figs.~\ref{fig:ejLunoinfcr}, \ref{fig:ejLu0204}, and~\ref{fig:ejLu02yz}
are not satisfactory.
Especially, the critical angular-momentum, $I_{\rm c}$,
where the one-wobbling-phonon excitation energy vanishes
and the main component of angular-momentum vector
in the intrinsic frame changes from that of the alignment-axis ($x$)
to the largest-inertia axis ($y$), is too small.
The experimental excitation energy does not vanish
in the observed range of angular momentum,
and therefore $I_{\rm c} > 91/2$ at least.
The calculated result presented so far, which satisfies this inequality,
is the case of the high-frequency cranking around the $x$ axis
with $\omega_x=0.4$, $\omega_y=\omega_z=0.01$ MeV/$\hbar$
displayed in the lower panels of Fig.~\ref{fig:ejLu0204};
the wobbling excitation energy is, however, too high in this case.
In order to study the properties of the electromagnetic transition probabilities,
the intrinsic nuclear shape, which is kept constant in the present work,
and the geometry of the angular-momentum vector in the intrinsic frame
are important.
Therefore, we slightly change the cranking frequencies and
make the critical angular-momentum higher still keeping
the one-wobbling-phonon excitation energy relatively low
as in the experimental data:
Figure~\ref{fig:ejLu02std} depicts the wobbling-excitation energies
and the expectation values of the angular-momentum vector for the cranked
mean-field with the frequencies,
$\omega_x=0.20$, $\omega_y=0.0$, $\omega_z=0.01$ MeV$/\hbar$.
The agreement of the excitation energy is much better,
although that of the two-phonon-wobbling band is still too high.
We found it difficult to obtain such low excitation energy
for the two-phonon wobbling band in the present calculation.

\begin{figure*}[!htb]
\begin{center}
\includegraphics[width=155mm]{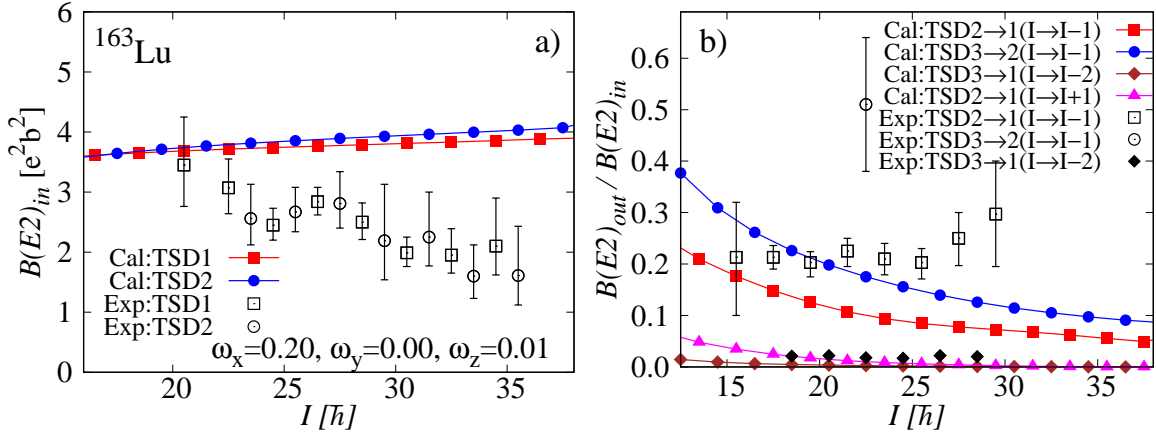}
\vspace*{-4mm}
\caption{(Color online)
The $I\rightarrow I-2$ in-band $E2$ transition probabilities (left panel) and
the $I\rightarrow I\pm 1$ out-of-band to in-band $B(E2)$ ratios (right panel)
are compared with the experimental data in $^{163}$Lu.
The cranked mean-field with the frequencies,
$\omega_x=0.20$, $\omega_y=0.0$, $\omega_z=0.01$ MeV$/\hbar$, are used
corresponding to Fig.~\ref{fig:ejLu02std}.
The experimental data are taken from~\cite{Jens02,Gorg04,Hama01}.
}
\label{fig:trLu02std}
\end{center}
\end{figure*}

\begin{figure}[!htb]
\begin{center}
\includegraphics[width=75mm]{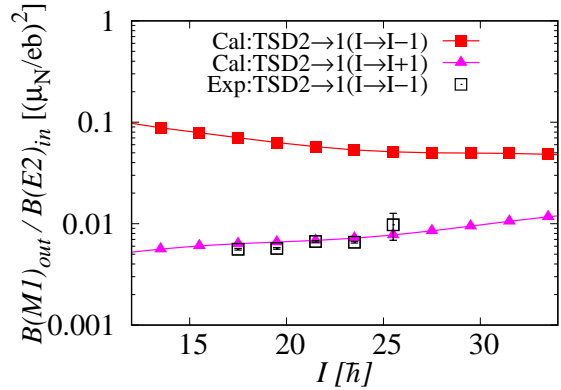}
\vspace*{-4mm}
\caption{(Color online)
The $I\rightarrow I\pm 1$ out-of-band $B(M1)$ to in-band $B(E2)$ ratios
are compared with the experimental data in $^{163}$Lu.
The cranked mean-field with the frequencies,
$\omega_x=0.20$, $\omega_y=0.0$, $\omega_z=0.01$ MeV$/\hbar$, are used
corresponding to Fig.~\ref{fig:ejLu02std}.
The experimental data are taken from~\cite{Hama01}.
}
\label{fig:tmLu02std}
\end{center}
\end{figure}

With this choice of the cranking frequencies and the excitation spectrum
of Fig.~\ref{fig:ejLu02std},
the calculated $I\rightarrow I-2$ in-band $E2$ transition probabilities
and the ratios of the $I\rightarrow I \pm 1$ out-of-band to in-band $B(E2)$
are compared with the experimental data in Fig.~\ref{fig:trLu02std}.
As it is seen, the in-band $B(E2)$ values are almost constant and agree
with the experimental value at low spin, $I\approx 20$.
However, the measured values decrease as a function of spin
in contrast to the calculation.
As for the out-of-band $B(E2)$ values from TSD2 to TSD1,
there are two possible transitions with $I \rightarrow I \pm 1$.
The rotor model predicts~\cite{BM75} that
the $I \rightarrow I-1$ transitions are much stronger for
the so-called ``positive-$\gamma$ rotation'' (see below for explanation)
in agreement with the projection calculation.
Only the $I \rightarrow I-1$ transitions are measured in experiment,
which is considered to be consistent with the rotor model prediction.
The calculated $I \rightarrow I-1$ $B(E2)_{\rm out}/B(E2)_{\rm in}$ ratio
from TSD2 to TSD1 at low spin, $I\approx 15$,
is also comparable with the experimentally measured value,
but the calculated ratios decrease as functions of spin,
which is in agreement with the rotor model prediction
because of the $1/I$ factor of
the relevant squared Clebsch-Gordan coefficients~\cite{BM75}.
In contrast, the measured values are almost constant or
even increase at the highest spins observed.
The observed$B(E2)_{\rm out}/B(E2)_{\rm in}$ ratio from TSD3 to TSD2
is almost factor two larger than that from TSD2 to TSD1,
but the calculated values for both transitions are considerably smaller
at higher spins, $I\gtsim 22$.
As for the $B(M1)$ transitions,
the calculated out-of-band $B(M1)$ to in-band $B(E2)$ ratio
is displayed in Fig.~\ref{fig:tmLu02std}.
The out-of-band $B(M1)$ transitions with $I \rightarrow I-1$
are also larger than those with $I \rightarrow I+1$,
but the calculated values of the $B(M1)$ ratio are
about one order of magnitude larger than the measured values:
The measured values of the $B(M1:I \rightarrow I-1)$
are of the same order of magnitude
as the calculated values for $I \rightarrow I+1$ transitions.
These results are very similar to those obtained in the previous
calculation~\cite{SS09}, where another microscopic approach,
the QRPA formalism, is employed.
It should be emphasized that in both the QRPA and the present projection method
the obtained results for $B(E2)$ strongly support the validity
of the basic picture of the simple triaxial-rotor model
(see also Ref.~\cite{TS16} for the similar conclusion).
Recently it was suggested~\cite{FD15} that the inclusion of
the isovector type schematic interaction composed of
the orbital angular-momentum resolves the difficulty of an order of magnitude
overestimation of the out-of-band $B(M1)$ values,
but we were not able to confirm it.  Further study of this point is needed.

The almost constant in-band $B(E2)$ values and the $1/I$-decreasing trend
of the out-of-band $B(E2)$ values are the result of
the fixed triaxial deformation and the fixed rotational axis
(see, e.g., Refs.~\cite{BM75,Mar79}).
The dominance of $I\rightarrow I-1$ transitions
for the out-of-band $B(E2)$ and $B(M1)$ values
is characteristic for the positive-$\gamma$ rotation, namely,
the rotation about the short axis,
($x$ axis with $0 < \gamma < 60^\circ$).
This feature is realized for the transverse wobbling.
If the nucleus rotates around the largest-inertia axis (medium axis)
of the core ($y$ axis with $0 < \gamma < 60^\circ$),
the $I\rightarrow I+1$ out-of-band transitions dominate
in the rotor model~\cite{BM75,Mar79}:
This feature is realized for simple wobbling of
the core (even-even nucleus)~\cite{TS16}
and also for the longitudinal wobbling.
Since the rotation-axis is quite often chosen to be the $x$ axis for
the study of the high-spin states,
such a rotation scheme around the largest inertia axis corresponds
to the triaxial deformation with $-60^\circ < \gamma < 0$, i.e.,
the ``negative-$\gamma$ rotation''.
Therefore, which out-of-band $B(E2)$ transition is stronger,
that with $I\rightarrow I+1$ or with $I\rightarrow I-1$,
is crucial to distinguish the positive-$\gamma$ or negative-$\gamma$ rotation,
which was first emphasized in Ref.~\cite{SM95}.
In the case of the odd-$A$ nucleus with a highly-alignable quasiparticle,
these positive-$\gamma$ and negative-$\gamma$ rotation just correspond
to the difference between the transverse and longitudinal wobbling,
respectively, and the observed data for $^{163}$Lu clearly indicate
transverse wobbling~\cite{FD14}.

\begin{figure*}[!htb]
\begin{center}
\includegraphics[width=155mm]{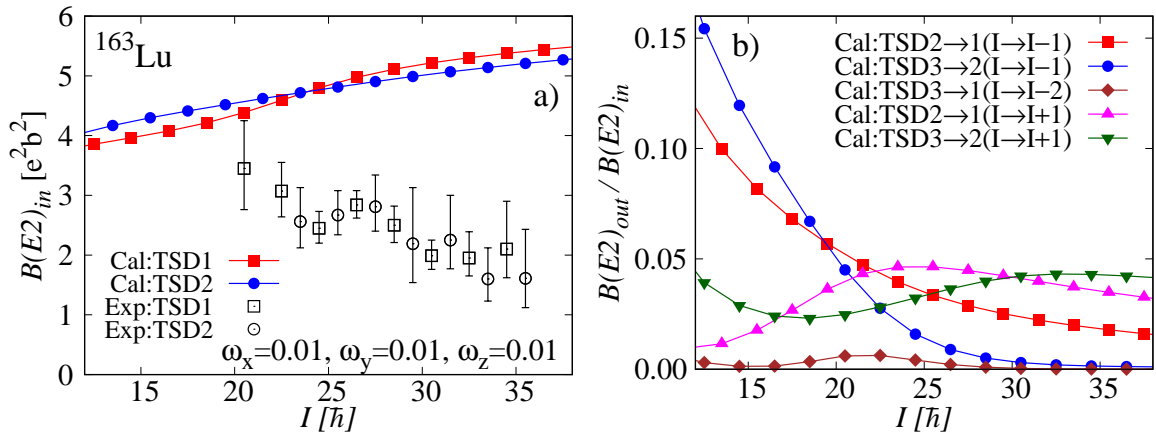}
\vspace*{-4mm}
\caption{(Color online)
The $I\rightarrow I-2$ in-band $E2$ transition probabilities (left panel)
and the $I\rightarrow I\pm 1$ out-of-band to in-band $B(E2)$ ratios
(right panel) in $^{163}$Lu.
The cranked mean-field with the frequencies,
$\omega_x=\omega_y=\omega_z=0.01$ MeV$/\hbar$, are used
corresponding to the lower panels of Fig.~\ref{fig:ejLunoinfcr}.
Note that the scale for the $B(E2)$ ratios is enlarged
from that in Fig.~\ref{fig:trLu02std}.
}
\label{fig:trLuinfcr}
\end{center}
\end{figure*}

In this way, the axis of rotation is also important for
the electromagnetic properties of the nuclear wobbling motion.
Although it does not correspond to the observed case in $^{163}$Lu,
we show in Fig.~\ref{fig:trLuinfcr}
the in-band $B(E2)$ values and the out-of-band to in-band $B(E2)$ ratios
for the case of the infinitesimally cranked mean-field with
$\omega_x=\omega_y=\omega_z=0.01$ MeV$/\hbar$, for which
the excitation spectrum and the expectation values of the angular-momentum vector
are shown in the lower panels of Fig.~\ref{fig:ejLunoinfcr}.
In this case the in-band $B(E2)$ values
in the left panel of Fig.~\ref{fig:trLuinfcr}
increase as a function of spin even if the deformation is kept constant.
This is because the main component of the angular-momentum vector
in the intrinsic frame changes from that of the $x$ axis to the $y$ axis
as it is shown in the lower-right panel of Fig.~\ref{fig:ejLunoinfcr}.
The in-band $B(E2)$ values are proportional to
$\bigl|\langle y^2-z^2 \rangle\bigr|^2$ when rotating around the $x$ axis,
while they are proportional to
$\bigl|\langle z^2-x^2 \bigr\rangle\bigr|^2 $when rotating around the $y$ axis.
The latter is larger for
the present triaxial deformation with $\gamma({\rm WS})=18^\circ$.
Therefore, the gradual change of the rotation axis from the $x$ axis
to the $y$ axis increases the in-band $B(E2)$ value
as it is displayed in the left panel of Fig.~\ref{fig:trLuinfcr}.
This effect is visible as very tiny effect in Fig.~\ref{fig:trLu02std}.
The in-band $B(E2)$ values only slightly
increase as a function of spin due to the gradual increase
of the component of the angular-momentum along the $y$ axis
in the right panel of Fig.~\ref{fig:ejLu02std}.
Although the absolute value of the out-of-band $B(E2)$
is too small in this case,
the effect of changing the rotation-axis is even more drastic
for the out-of-band $B(E2)$
as it is shown in the right panel of Fig.~\ref{fig:trLuinfcr}.
The dominance of the $I\rightarrow I-1$ or
of the $I\rightarrow I+1$ transitions
exchanges near the critical angular-momentum of
the vanishing one-phonon energy, $I_{\rm c}\approx 22$,
in the lower panel of Fig.~\ref{fig:ejLunoinfcr}.
The $I\rightarrow I+1$ out-of-band $B(E2)$ values of both
the transitions from TSD2 to TSD1 and from TSD3 to TSD2 are much larger
for $I \gtsim I_{\rm c}$ in contrast to the opposite feature at low spins.

\subsection{Results with larger $\bm{\gamma}$ deformation}
\label{sec:largegam}

Until now we have used the triaxial deformation
in Eq.~(\ref{eq:sgammaden}), which is the selfconsistent value for
the Nilsson or Woods-Saxon Strutinsky method.
The triaxiality parameter also affects the wobbling motion,
especially the $B(E2)$ values~\cite{BM75}.
Therefore, we show here some results of the projection calculation
with larger triaxial deformation for the mean-field.
We choose rather arbitrarily $\gamma({\rm WS})=30^\circ$
keeping the other deformation parameters $\beta_2=0.42$ and $\beta_4=0.02$.
It is noted again that $\gamma({\rm WS})=30^\circ$ of
the Woods-Saxon potential corresponds to smaller triaxial deformation
of the density distribution; for $\beta_2\sim 0.42$,
\begin{equation}
 \gamma({\rm WS})\sim 30^\circ
 \quad\Leftrightarrow\quad
 \gamma({\rm den})\sim 21-22^\circ .
\label{eq:lgammaden}
\end{equation}
It should be mentioned that the positive parity proton orbit at the Fermi
surface is not the one originating from the $i_{13/2}$ state for
$(\beta_2,\beta_4,\gamma({\rm WS}))=(0.42,0.02,30^\circ)$, which is crossed
at $\gamma({\rm WS})\approx 25^\circ$ by a orbit
which is mainly of $N_{\rm osc}=4$ (see Fig.~1 of Ref.~\cite{SS09}).
We select to occupy the second excited quasiparticle
near the Fermi surface originating from the $i_{13/2}$ particle
for an odd proton, which is necessary to realize the TSD states.

\begin{figure*}[!htb]
\begin{center}
\includegraphics[width=155mm]{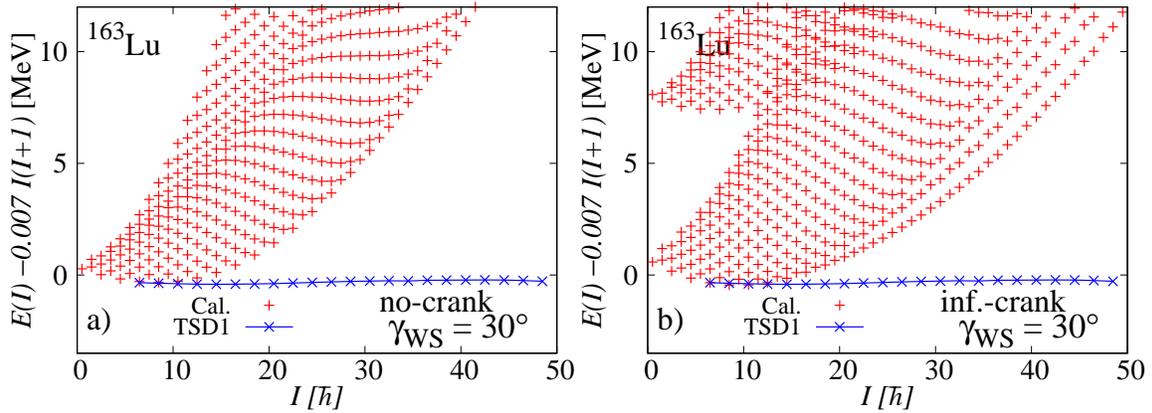}
\vspace*{-4mm}
\caption{(Color online)
Same as Fig.~\ref{fig:sLunoinfcr} but the result of projection
from the mean-field with larger triaxial deformation
$\gamma({\rm WS})=30^\circ$.
}
\label{fig:sLu30gnoinfcr}
\end{center}
\end{figure*}

In Fig.~\ref{fig:sLu30gnoinfcr} we show the calculated spectrum
with the non-cranked mean-field (left panel)
and with the infinitesimally cranked mean-field
with $\omega_x=\omega_y=\omega_z=0.01$ MeV$/\hbar$ (right panel)
just like in Fig.~\ref{fig:sLunoinfcr}.
As in the case of $\gamma({\rm WS})=18^\circ$ the multiple band structure
characteristic for the wobbling motion appears.
Compared with Fig.~\ref{fig:sLunoinfcr},
the moments of inertia for the rotational bands are slightly smaller
in Fig.~\ref{fig:sLu30gnoinfcr} than in Fig.~\ref{fig:sLunoinfcr}.
This can be naturally understood.  As already studied in the previous
section, the main rotation-axis is the short ($x$) axis due to the presence
of the aligned $\pi i_{13/2}$ particle, and the core moment of inertia around
this axis decreases as a function of $\gamma$; see Fig.~\ref{fig:momi}.
However, the value of the cranking inertia of the $x$ axis
does not change so much at $\gamma({\rm WS})=30^\circ$ compared
with that at $\gamma({\rm WS})=18^\circ$
because of the bump-like behavior in Fig.~\ref{fig:momi}.
Therefore the moments of inertia of the wobbling bands are
only slightly reduced in Fig.~\ref{fig:sLu30gnoinfcr}
in comparison with those in Fig.~\ref{fig:sLunoinfcr}.

\begin{figure*}[!htb]
\begin{center}
\includegraphics[width=155mm]{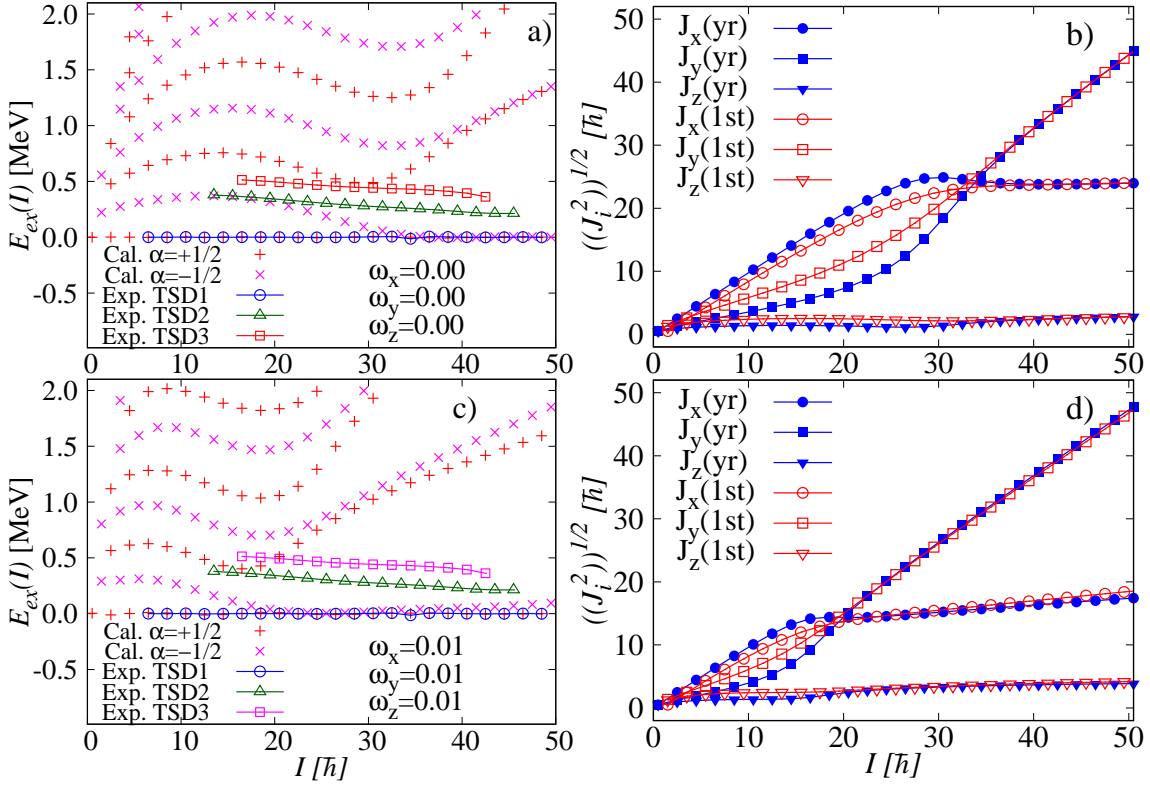}
\vspace*{-4mm}
\caption{(Color online)
Same as Fig.~\ref{fig:ejLunoinfcr} but the result of projection
from the mean-field with larger triaxial deformation
$\gamma({\rm WS})=30^\circ$.
}
\label{fig:ejLu30gnoinfcr}
\end{center}
\end{figure*}

The excitation spectrum and the expectation values of the angular-momentum
components in the intrinsic frame are displayed
in the upper and lower panels of Fig.~\ref{fig:ejLu30gnoinfcr}
for the cases of the non-cranked and
of the infinitesimally cranked mean-fields, respectively.
By comparing Figs.~\ref{fig:ejLunoinfcr} and~\ref{fig:ejLu30gnoinfcr},
it can be seen that the excitation energy of the wobbling phonon
is smaller for the mean-field with larger triaxiality,
$\gamma({\rm WS})=30^\circ$,
i.e., the wobbling-phonon excitation energy
decreases when the triaxiality increases.
Because of this the critical angular-momentum of the vanishing one-phonon
excitation energy is shifted to lower spins in Fig.~\ref{fig:ejLu30gnoinfcr}.
It is worth mentioning that
in the case of larger triaxial deformation of Eq.~(\ref{eq:lgammaden})
the excitation spectrum after the critical frequency is different from the one
in the case of smaller triaxial deformation of Eq.~(\ref{eq:sgammaden}).
The signature partner bands with $\alpha=\pm \frac{1}{2}$ are
almost degenerate after the critical spin in the this case, i.e.,
there is one ${\mit\Delta}I=1$ band instead of two ${\mit\Delta}I=2$ bands,
for $I \gtsim 32$ in the upper panels of Fig.~\ref{fig:ejLu30gnoinfcr}
and for $I \gtsim 20$ in the lower panels of Fig.~\ref{fig:ejLu30gnoinfcr},
while the signature splittings are significant in Fig.~\ref{fig:ejLunoinfcr},
even though the behavior of the expectation values of angular-momentum vectors
are rather similar in Figs.~\ref{fig:ejLu30gnoinfcr} and~\ref{fig:ejLunoinfcr}.
The ``signature quantum number'' is severely broken
in the case of larger triaxial deformation in Fig.~\ref{fig:ejLu30gnoinfcr}.
We think that the reason is the following.  In the present case the component
of angular-momentum along the long axis ($z$ axis) is always small,
so that the rotational axis lies in the $xy$-plane.
If the both the $x$ and $y$ components are sizable
the symmetry with respect to the $180^\circ$-rotation
around the rotation axis is broken.
However, if the mean-field is axially symmetric about the $z$ axis,
the signature symmetry is still present.
In case of the smaller triaxial deformation in Eq.~(\ref{eq:sgammaden})
($\gamma({\rm den})\approx 11-12^\circ$)
the signature symmetry is not so strongly broken and
sizable signature splitting appears for the wobbling bands
based on the highly-alignable $\pi i_{13/2}$ particle.
In contrast, the triaxial deformation in Eq.~(\ref{eq:lgammaden})
($\gamma({\rm den})\approx 21-22^\circ$) is large enough to strongly break
the signature symmetry.  The signature splitting is getting sizable
at highest spins displayed in the lower panels of Fig.~\ref{fig:ejLu30gnoinfcr},
because the $x$-component is getting smaller compared to the $y$-component
and the nucleus rotates mainly around the principal $y$ axis,
which makes the signature an approximately good quantum-number again.
Thus, these are interesting examples of the interplay
of the dynamical motion of the angular-momentum vector and
the triaxial deformed mean-field in the intrinsic frame.

\begin{figure*}[!htb]
\begin{center}
\includegraphics[width=155mm]{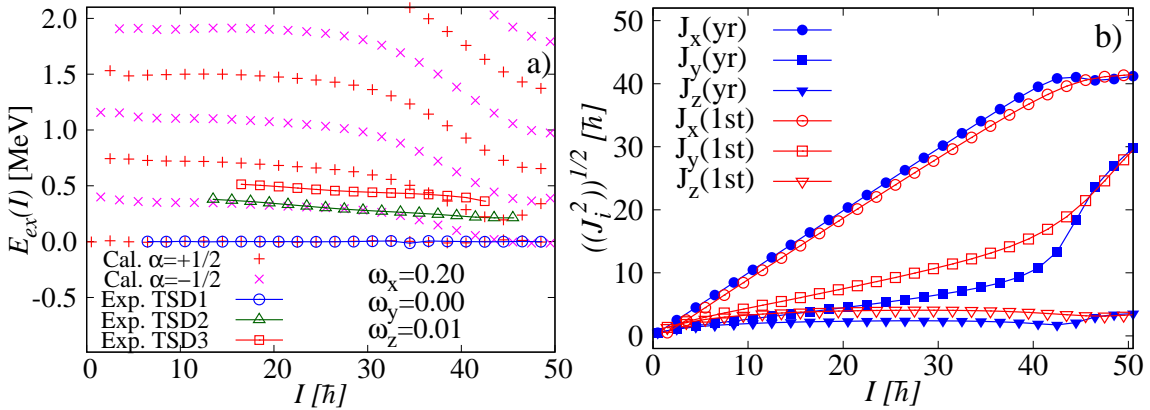}
\vspace*{-4mm}
\caption{(Color online)
Same as Fig.~\ref{fig:ejLu02std} but the result of projection
from the mean-field with larger triaxial deformation
$\gamma({\rm WS})=30^\circ$.
}
\label{fig:ejLu30g02std}
\end{center}
\end{figure*}

\begin{figure*}[!htb]
\begin{center}
\includegraphics[width=155mm]{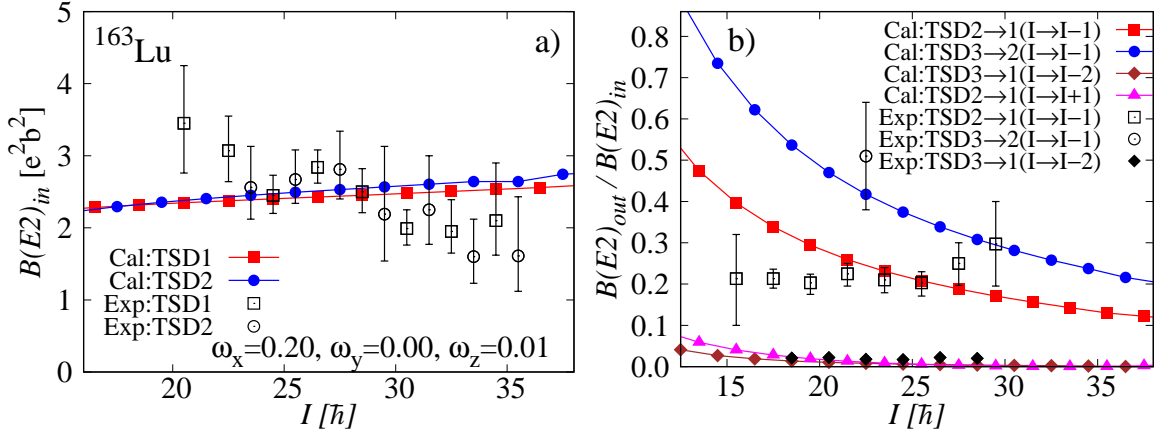}
\vspace*{-4mm}
\caption{(Color online)
Same as Fig.~\ref{fig:trLu02std} but the result of projection
from the mean-field with larger triaxial deformation
$\gamma({\rm WS})=30^\circ$.
}
\label{fig:trLu30g02std}
\end{center}
\end{figure*}

It is well-known that the effect of triaxial deformation is
more important for the $E2$ transition probabilities~\cite{BM75}.
As in the previous case of smaller triaxial deformation, we present,
for the case of larger triaxiality $\gamma({\rm WS})=30^\circ$,
the result of projection from the cranked mean-field with
$\omega_x=0.20$, $\omega_y=0.0$, $\omega_z=0.01$ MeV$/\hbar$,
in which the value of the critical angular-momentum is relatively large
and still the one-phonon excitation energy is relatively low.
We first show in Fig.~\ref{fig:ejLu30g02std}
the excitation spectrum and the expectation values of
the angular-momentum vector in the intrinsic frame,
just like in Fig.~\ref{fig:ejLu02std},
where the triaxiality is smaller, $\gamma({\rm WS})=18^\circ$.
As in the case of the non-cranked or of the infinitesimally cranked mean-field,
the excitation energies are smaller and the agreement of the TSD1 excitation
energy is better, although the calculated TSD2 excitation energies are still
higher than the experimental data.
The calculated in-band $B(E2)$ values and
the out-of-band to in-band $B(E2)$ ratios
are compared with the experimental data in Fig.~\ref{fig:trLu30g02std}.
Comparing in-band $B(E2)$ values for the cases with $\gamma({\rm WS})=18^\circ$
and $30^\circ$ in the left panels
of Figs.~\ref{fig:trLu02std} and~\ref{fig:trLu30g02std},
the latter is considerably smaller.
This can be understood by the rotor model~\cite{BM75}.
As the rotation-axis is mainly the $x$ axis in both cases
(see the right panels of
Figs.~\ref{fig:ejLu02std} and~\ref{fig:ejLu30g02std}),
the $B(E2)$ values are proportional to
$\bigl|\langle y^2-z^2 \rangle\bigr|^2 \propto \cos^2{(\gamma+30^\circ)}$,
which is a decreasing function of $\gamma$ for $0 < \gamma < 60^\circ$.
On the other hand, the out-of-band to in-band $B(E2)$ ratios
are considerably larger for larger $\gamma$ deformation,
which can be also understood by the rotor model~\cite{BM75}.
The large average value of $B(E2)_{\rm out}/B(E2)_{\rm in}$ ratios
for the TSD2 to TSD1 transitions,
which is considered to be crucial to identify the wobbling motion,
is better described by the calculation
with the larger $\gamma$ value in Eq.~(\ref{eq:lgammaden}).
The calculated $B(M1)_{\rm out}/B(E2)_{\rm in}$ ratios
are about factor two lager for the $I \rightarrow I-1$ transitions
than for $\gamma({\rm WS})=18^\circ$ (not shown),
while those for the $I \rightarrow I+1$ transitions are similar;
i.e., the overestimation of the $B(M1)$ values is a little bit more serious.

The experimental in-band $B(E2)$ values decrease as a function of spin,
and $B(E2)_{\rm out}/B(E2)_{\rm in}$ ratios are almost constant
or even increase at the highest spins.
It is very difficult to reproduce these trends in the calculation
as long as constant deformation is assumed.
It is discussed~\cite{SS09} that
a considerable increase of the $\gamma$ deformation
as a function of spin can explain these features
of the in-band and out-of-band $B(E2)$ values,
although the selfconsistent mean-field calculations suggest
that the deformation does not change so much.
In order to perform the angular-momentum-projection calculation
for such mean-field changing with spin,
one has to prepare several mean-fields with different triaxiality $\gamma$
and employ the configuration-mixing calculation like in Ref.~\cite{TS16},
where no cranking is performed.
Such calculations combined with the finite cranking frequencies
are interesting but out of scope in the present investigation.

\subsection{Example for the case of the odd proton in a non-intruder orbit}
\label{sec:xpodd}

For the sake of completeness we briefly discuss
the spectrum obtained by projection from the mean-field state,
where the odd proton in $^{163}$Lu does not occupy
the high-$j$ intruder orbit $i_{13/2}$.
As it is mentioned in the previous subsection, the lowest proton orbit
near the Fermi surface at $\gamma({\rm WS})=30^\circ$
originates mainly from $N_{\rm osc}=4$
(see Fig.~1 of Ref.~\cite{SS09}).
This relatively low-$j$ orbit is now occupied by the odd proton quasiparticle
to generate the mean-field state,
from which the projection calculation is performed.

\begin{figure*}[!htb]
\begin{center}
\includegraphics[width=155mm]{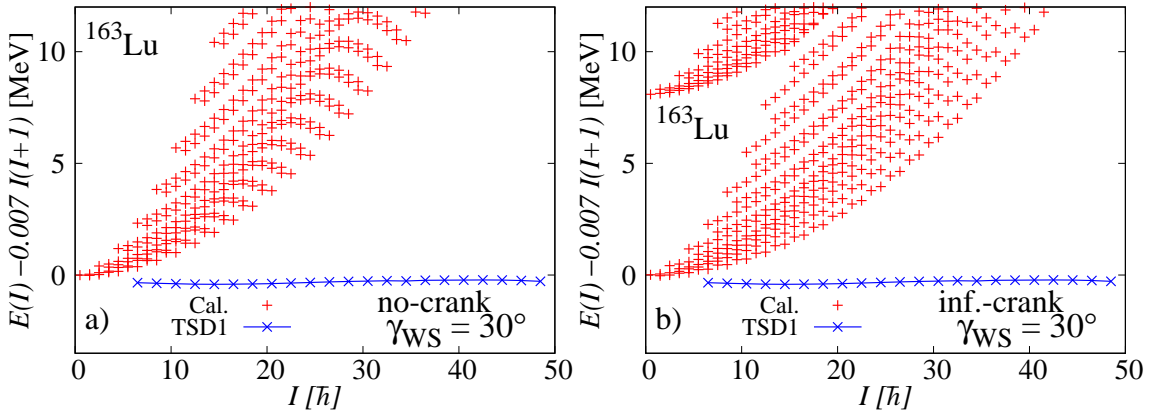}
\vspace*{-4mm}
\caption{(Color online)
Same as Fig.~\ref{fig:sLu30gnoinfcr} but
the odd-proton particle occupies low-$j$ orbit in this calculation.
}
\label{fig:sLu30gxpnoinfcr}
\end{center}
\end{figure*}

Figure~\ref{fig:sLu30gxpnoinfcr} displays the calculated spectrum
with the non-cranked mean-field (left panel)
and with the infinitesimally cranked mean-field
with $\omega_x=\omega_y=\omega_z=0.01$ MeV$/\hbar$ (right panel)
just like in Figs.~\ref{fig:sLunoinfcr} and~\ref{fig:sLu30gnoinfcr}.
The multiple-band structure emerges in both cases as in the previous cases.
However, apparently the slopes of the wobbling bands are steeper than
those in Fig.~\ref{fig:sLu30gnoinfcr};
namely their calculated moments of inertia are even smaller than the case
with the high-$j$ orbit being occupied.
The lowest energy state has $I^\pi=1/2^+$ for the case of no-cranking,
and $I^\pi=3/2^+$ for the case of infinitesimal cranking, respectively.

\begin{figure*}[!htb]
\begin{center}
\includegraphics[width=155mm]{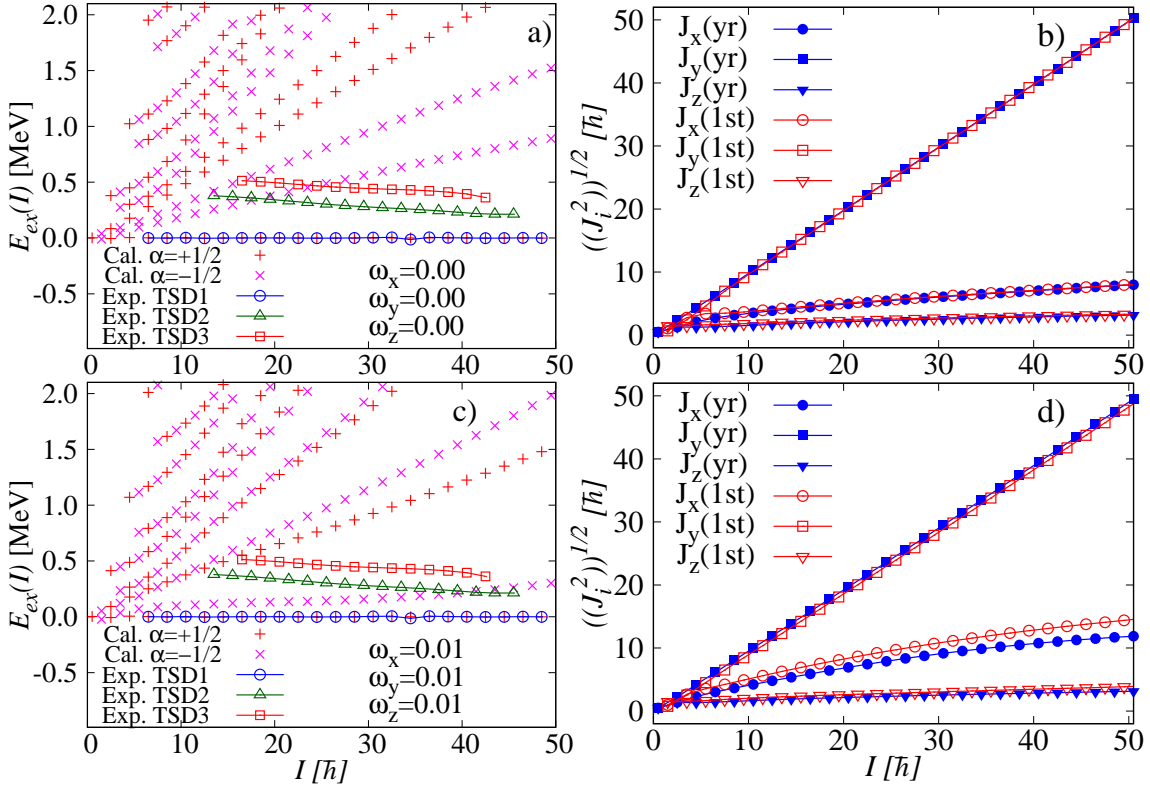}
\vspace*{-4mm}
\caption{(Color online)
Same as Fig.~\ref{fig:ejLu30gnoinfcr} but
the odd-proton particle occupies low-$j$ orbit in this calculation.
}
\label{fig:ejLu30gxpnoinfcr}
\end{center}
\end{figure*}

The excitation energies and the expectation values of the angular-momentum
components in the intrinsic frame are displayed
in the upper and lower panels of Fig.~\ref{fig:ejLu30gxpnoinfcr}
for the cases of the non-cranked and of the infinitesimally cranked mean-field,
respectively.
The dominant component of the angular-momentum is always along
the largest inertia axis ($y$ axis) in both cases, and, consequently,
the excitation energies increase monotonically as functions of spin,
just like for the even-even core nucleus.
The odd-proton aligns its angular-momentum vector mainly along the $y$ axis,
although the alignment is rather small, and this case roughly corresponds
to the longitudinal wobbling.
In this way, the basic picture of the transverse wobbling
and of the longitudinal wobbling,
proposed in Ref.~\cite{FD14} and discussed in Sec.~\ref{sec:transwob},
is justified by our fully microscopic calculations
in the framework of the angular-momentum-projection method.
It is interesting to mention that
the signatures of the yrast and excited bands are
$\alpha=+1/2,\,-1/2,\,-1/2,\,+1/2,\,+1/2,\cdots$
in the result with the non-cranked mean-field
in the upper-left panel of Fig.~\ref{fig:ejLu30gxpnoinfcr},
which is different from the simple alternating pattern in the case
of occupying the high-$j$ intruder orbit.
This is because the non-intruder orbit of the odd proton
is a strongly mixed state of the spherical shell-model orbits,
$s_{1/2}$, $d_{3/2}$, $d_{5/2}$ and $g_{7/2}$,
and its angular momentum $j$ is not definite.
Consequently, the coupling scheme between the angular momenta
of the odd proton and the even-even core
is not so simple as in the case of the high-$j$ intruder orbit.
Moreover, the excitation energy
of the second excited band is smaller than twice
the excitation energy of the first excited band.
In contrast, the alternating pattern of the signatures is recovered
with the infinitesimal cranking
in the lower-left panel of Fig.~\ref{fig:ejLu30gxpnoinfcr},
and the second wobbling energy is considerably larger than twice
the first wobbling energy.
At present we do not see a clear reason why this kind of qualitative
difference appears as a result of the infinitesimal cranking.

\subsection{Result with Gogny D1S interaction}
\label{sec:gogny}

It is known that the density-dependent term
in the Gogny or Skyrme interactions causes fundamental problems for
beyond mean-field calculations including the angular-momentum-projection
(see, e.g., Refs.~\cite{DSN07,Rob10} and references therein).
The problem seems to be more serious for odd and odd-odd nuclei
than for even-even nuclei.
Although infinitesimal cranking around all three principal axes
has been applied in the even-even nucleus $^{164}$Er in Ref.~\cite{TS16},
we found for $^{163}$Lu that infinitesimal cranking
with respect to more than one axis suffers from these problems.
We were able to obtain reasonable result of the projection calculation
for the mean-field with only one-dimensional cranking.
Therefore we only show the result of such a case.
All the calculations have been done by expanding the HFB states
within the isotropic harmonic oscillator basis as for the calculation
of the Woods-Saxon mean-field.  We have used the same model space
composed of the basis states with the maximum oscillator shells up to
$N_{\rm osc}^{\rm max}=12$.

\begin{table*}[!htb]
\begin{center}
\begin{tabular}{cccccccccc}
\hline \hline
 $\omega_x$[MeV] & $\langle J_x \rangle$[$\hbar$] &
 ${\widebar{\Delta}}_{\rm n}$[MeV] & ${\widebar{\Delta}}_{\rm p}$[MeV]
 & $\widebar{R}$[fm] &
 $\beta_2({\rm den})$ & $\gamma({\rm den})$ &
 $\alpha_{40}({\rm den})$ & $\alpha_{42}({\rm den})$ &
 $\alpha_{44}({\rm den})$ \\
\hline
$0.050$ & $ 8.1$  & $0.776 $ & $0.496 $ & $6.880$ & $0.443$ & $10.7^\circ$ & $0.162$ & $-0.0394$ & $-0.00386$ \\
$0.100$ & $11.1$  & $0.753 $ & $0.470 $ & $6.880$ & $0.443$ & $10.9^\circ$ & $0.163$ & $-0.0406$ & $-0.00373$ \\
$0.150$ & $14.1$  & $0.722 $ & $0.416 $ & $6.880$ & $0.443$ & $11.0^\circ$ & $0.162$ & $-0.0416$ & $-0.00359$ \\
$0.200$ & $17.3$  & $0.677 $ & $0.322 $ & $6.880$ & $0.443$ & $11.2^\circ$ & $0.160$ & $-0.0429$ & $-0.00342$ \\
$0.250$ & $20.8$  & $0.610 $ & $0.0867$ & $6.878$ & $0.442$ & $11.4^\circ$ & $0.157$ & $-0.0449$ & $-0.00320$ \\
$0.300$ & $24.4$  & $0.515 $ & $0.000 $ & $6.877$ & $0.440$ & $11.7^\circ$ & $0.156$ & $-0.0470$ & $-0.00305$ \\
$0.350$ & $28.0$  & $0.373 $ & $0.000 $ & $6.874$ & $0.438$ & $12.0^\circ$ & $0.155$ & $-0.0494$ & $-0.00310$ \\
$0.400$ & $31.7$  & $0.0536$ & $0.000 $ & $6.870$ & $0.434$ & $12.3^\circ$ & $0.154$ & $-0.0520$ & $-0.00366$ \\
\hline \hline
\end{tabular}
\vspace*{4mm}
\caption{
The expectation value $\langle J_x \rangle$,
the neutron and proton average pairing gaps in Eq.~(\ref{eq:gavgap}),
nuclear radius and various non-zero
deformation parameters in Eq.~(\ref{eq:defparam}) with $\lambda \le 4$
as functions of the rotational frequency with respect to the $x$ axis,
$\omega_x$, of mean-field obtained by the cranked Gogny-HFB calculation
for the TSD yrast states of $^{163}$Lu.
}
\label{tab:wobgmf}
\end{center}
\end{table*}

In this Gogny-HFB approach the mean-field potential is generated by
the selfconsistent HFB procedure.
Since the $\pi i_{13/2}$ particle favors to align its angular-momentum
along the short ($x$) axis, we have performed the Gogny HFB calculation
by blocking the lowest positive-parity quasiproton with $x$ axis cranking.
With finite rotational frequencies we have found essentially
the same TSD mean-field parameters as obtained by the Nilsson-Strutinsky
calculation (e.g., in Ref.~\cite{BR04}; note the relation of
the triaxiality parameter in Eq.~(\ref{eq:sgammaden})).
We were not able to obtain a convergent solution
near zero rotational frequency.
The average pairing gaps in Eq.~(\ref{eq:gavgap})
and deformation parameters in Eq.~(\ref{eq:defparam})
of the HFB mean-field are tabulated in Table~\ref{tab:wobgmf}
as functions of the rotational frequency $\omega_x$,
where the expectation value of the angular-momentum $\langle J_x \rangle$
is also included.
Clearly there exists aligned angular-momentum along the $x$ axis
of the odd $i_{13/2}$-proton.
Note that for one-dimensional cranking around the $x$ axis
all non-zero deformation parameters are real.
Here the parameters $(\beta_2,\gamma)$ are determined, as usual, by
$\beta_2({\rm den})\equiv
\bigl[\alpha_{20}({\rm den})^2+2\alpha_{22}({\rm den})^2\bigr]^{1/2}$
and $\gamma({\rm den})\equiv
\tan^{-1}\bigl[-\sqrt{2}\alpha_{22}({\rm den})/\alpha_{20}({\rm den})\bigr]$,
which is exactly the same as in Eq.~(\ref{eq:gammaden}).
The average pairing gaps at the lowest frequency, $\omega_x=0.05$ MeV/$\hbar$,
for both neutrons and protons are smaller than
the even-odd mass differences of the ground states
in the neighboring even-even nuclei.
The proton gap is especially small because of the blocking effect.
The proton (neutron) pairing-correlations vanish
for $\omega_x \gtsim 0.25$ ($\omega_x \gtsim 0.40$) MeV/$\hbar$
as it is shown in Table~\ref{tab:wobgmf}.
The obtained triaxiality parameter $\gamma({\rm den})$ just corresponds
to the value in Eq.~(\ref{eq:sgammaden}),
and it is almost constant
or only slightly increases as the frequency $\omega_x$ increases,
which is consistent with the result of Ref.~\cite{FD15}
considering the different definition of the triaxiality parameter
as it is mentioned in Eqs.~(\ref{eq:sgammaden}) and~(\ref{eq:lgammaden});
see Ref.~\cite{SSM08} for the relation between
$(\beta_2({\rm WS}),\gamma({\rm WS}))$
(or $(\beta_2({\rm Nils}),\gamma({\rm Nils})))$
and $(\beta_2({\rm den}),\gamma({\rm den}))$.

\begin{figure*}[!htb]
\begin{center}
\includegraphics[width=155mm]{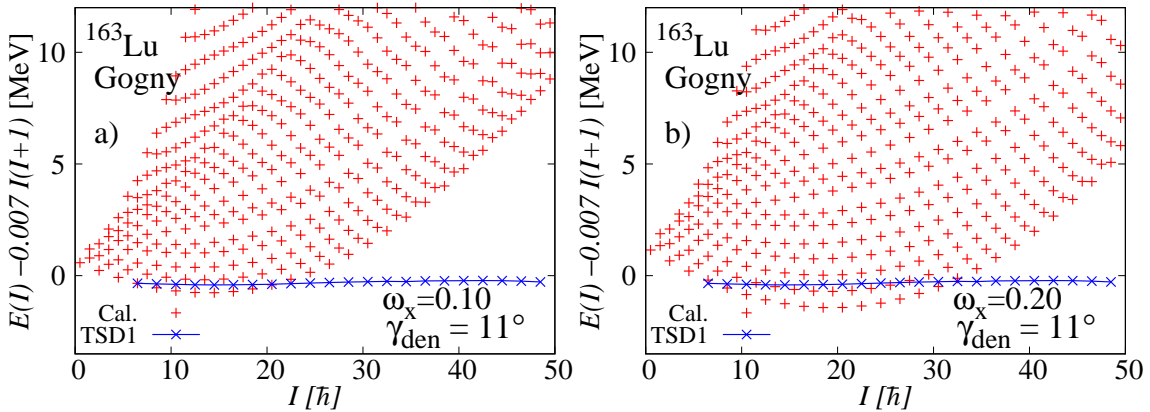}
\vspace*{-4mm}
\caption{(Color online)
Wobbling spectrum for $^{163}$Lu calculated by
the angular-momentum-projection method
with the Gogny D1S effective interaction.
The cranked mean-field with
$\omega_x=0.10$ ($\omega_x=0.20$), $\omega_y=\omega_z=0.0$ MeV$/\hbar$
is employed in the left (right) panel.
The energy of the experimental TSD1~\cite{Jens02} is also included
in each panel.
}
\label{fig:sLuGog0102}
\end{center}
\end{figure*}

\begin{figure*}[!htb]
\begin{center}
\includegraphics[width=155mm]{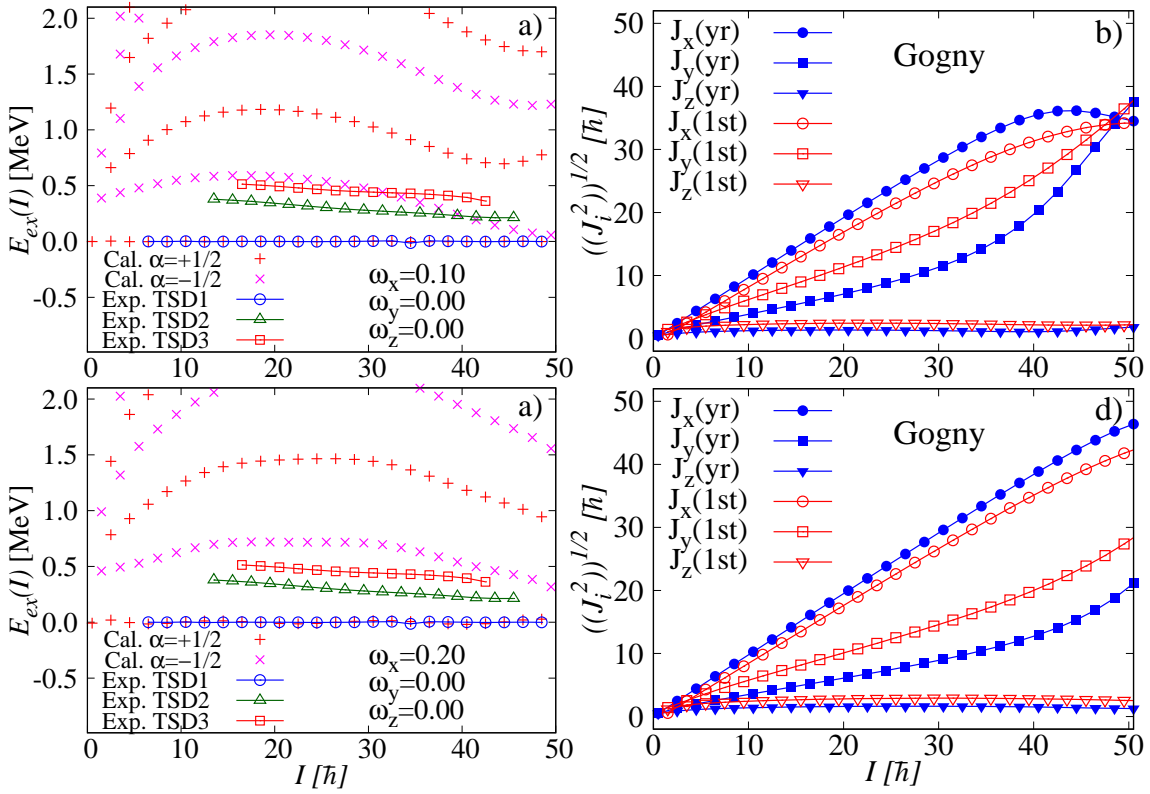}
\vspace*{-4mm}
\caption{(Color online)
The calculated excitation energy (left panels) and
the expectation values of the angular-momentum vector (right panels)
with using the Gogny D1S interaction in $^{163}$Lu.
The mean-field is cranked with the frequencies,
$\omega_x=0.10$, $\omega_y=\omega_z=0.0$ MeV$/\hbar$ (upper panels),
and
$\omega_x=0.20$, $\omega_y=\omega_z=0.0$ MeV$/\hbar$ (lower panels),
The experimental excitation energies of TSD1, TSD2, and TSD3~\cite{Jens02}
are also included in the left panels.
}
\label{fig:ejLuGog0102}
\end{center}
\end{figure*}

Figure~\ref{fig:sLuGog0102} shows the calculated spectrum with using
the Gogny D1S interaction, where the mean-field is cranked with
the rotational frequencies $\omega_x=0.10$ ($\omega_x=0.20$),
$\omega_y=\omega_z=0.0$ MeV$/\hbar$ in the left (right) panel
(see the mean-field parameters in Table~\ref{tab:wobgmf}).
The multiple-band structure, which is characteristic for
the nuclear wobbling motion, emerges also in this case.
Compared with the result of the Woods-Saxon mean-field and
the schematic interaction in Fig.~\ref{fig:sLunoinfcr}
or in Fig.~\ref{fig:sLuss02},
the slopes of the wobbling bands are less steep.
Namely, the moments of inertia are larger in the result
with the Gogny D1S interaction;
${\cal J}^{(1)}\approx 51 \hbar^2/$MeV at $I\approx 20$
for the calculation with $\omega_x=0.10$ MeV/$\hbar$ (the left panel),
and ${\cal J}^{(1)}\approx 65 \hbar^2/$MeV
with $\omega_x=0.20$ MeV/$\hbar$ (the right panel).

The excitation energy and the expectation values
of the angular-momentum vector in the intrinsic frame
are displayed in the upper and lower panels of Fig.~\ref{fig:ejLuGog0102},
for the case of the cranking frequency $\omega_x=0.10$ and 0.20 MeV/$\hbar$,
respectively.  Clearly the phonon excitations can also be seen for
the calculation with using the Gogny D1S interaction,
where the excitation energy first increases and then decreases
showing the characteristic behavior of transverse wobbling.
Apparently the resultant excitation spectrum is very similar to the one
in the case of the Woods-Saxon mean-field and the schematic interaction.
The critical point of vanishing one-phonon excitation energy just
corresponds to the point where the main component of the expectation values
of the angular-momentum vector changes from the $x$ axis to the $y$ axis
as it is shown in the right panels in Fig.~\ref{fig:ejLuGog0102}.
The excitation energy increases when the cranking frequency $\omega_x$
is increased in the same way as in the case of the Woods-Saxon mean-field.

\begin{figure*}[!htb]
\begin{center}
\includegraphics[width=155mm]{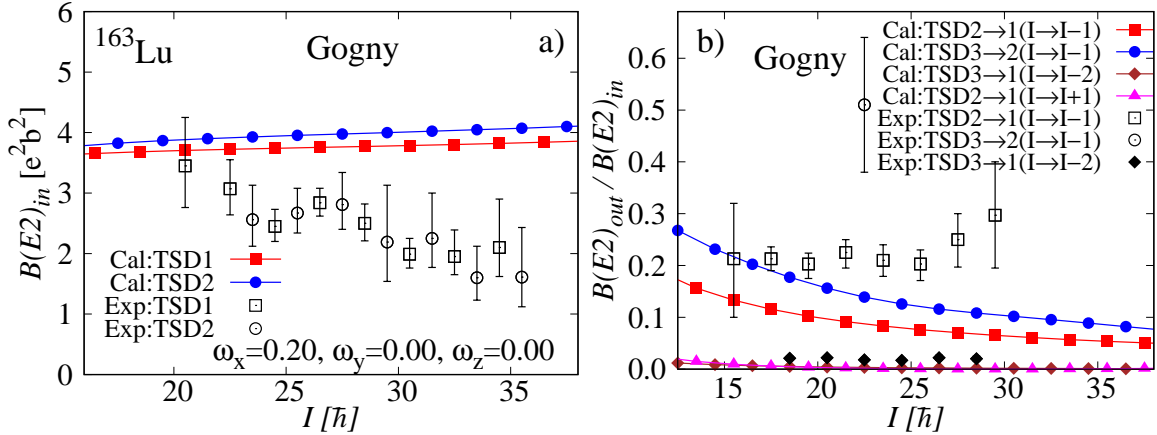}
\vspace*{-4mm}
\caption{(Color online)
The $I\rightarrow I-2$ in-band $E2$ transition probabilities (left panel) and
the $I\rightarrow I\pm 1$ out-of-band to in-band $B(E2)$ ratios (right panel),
which are calculated with the Gogny D1S interaction,
are compared with the experimental data in $^{163}$Lu.
The cranked mean-field with the frequencies,
$\omega_x=0.20$, $\omega_y=\omega_z=0.0$ MeV$/\hbar$, are used
corresponding to the lower panels of Fig.~\ref{fig:ejLuGog0102}.
}
\label{fig:trLuGog02std}
\end{center}
\end{figure*}

\begin{figure}[!htb]
\begin{center}
\includegraphics[width=75mm]{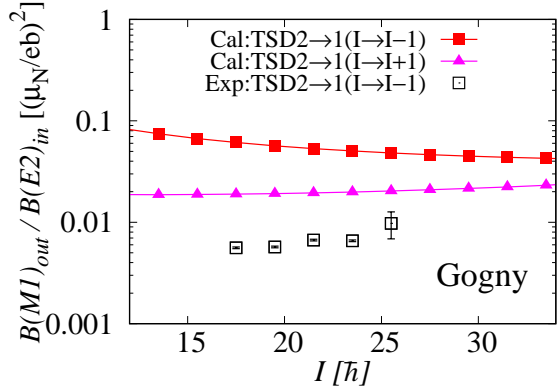}
\vspace*{-4mm}
\caption{(Color online)
The $I\rightarrow I\pm 1$ out-of-band $B(M1)$ to in-band $B(E2)$ ratios
calculated with the Gogny D1S interaction,
are compared with the experimental data in $^{163}$Lu.
The cranked mean-field with the frequencies,
$\omega_x=0.20$, $\omega_y=\omega_z=0.0$ MeV$/\hbar$, are used
corresponding to the lower panels of Fig.~\ref{fig:ejLuGog0102}.
}
\label{fig:tmLuGog02std}
\end{center}
\end{figure}

We show the in-band $B(E2)$ values and the out-of-band to in-band $B(E2)$
and $B(M1)$ ratios for the calculation using the Gogny D1S interaction
in Figs.~\ref{fig:trLuGog02std} and~\ref{fig:tmLuGog02std},
which are calculated for the mean-field
with $\omega_x=0.20$, $\omega_y=\omega_z=0.0$ MeV/$\hbar$,
corresponding to the excitation spectrum
in the lower panels of Fig.~\ref{fig:ejLuGog0102}.
The in-band $B(E2)$ values and the $B(E2)_{\rm out}/B(E2)_{\rm in}$ ratios
are very similar to
the ones of the Woods-Saxon mean-field in Figs.~\ref{fig:trLu02std}.
The $B(M1)_{\rm out}/B(E2)_{\rm in}$ ratios are also similar,
although the ratio with $I\rightarrow I-1$ transition are slightly smaller
and that with $I\rightarrow I+1$ transition are larger
than those in Fig.~\ref{fig:tmLu02std}.
All the characteristic features of the excitation energy spectrum and
of the transition probabilities are the same as in the
case of the Woods-Saxon mean-field and the schematic interaction.
If the same cranking frequencies are used
in the Woods-Saxon mean-field, i.e.,
$\omega_x=0.20$, $\omega_y=\omega_z=0.0$ MeV$/\hbar$,
the results of the Gogny-HFB and the Woods-Saxon mean-field
are much more similar: The only difference is that
the absolute value of the moment of inertia
for the wobbling band is larger in the calculation with the Gogny interaction.
This result clearly tells us that the wobbling motion calculated
by the microscopic angular-momentum-projection method does not
essentially depend on the details of the used effective interaction.
Therefore, we confirmed that the rotor-model picture of the wobbling motion
is validated by our microscopic projection calculations.

\subsection{Multi-cranked configuration-mixing with Gogny D1S interaction}
\label{sec:mccmgog}

Recently we have proposed the method of projected multi-cranked
configuration-mixing for a reliable description of
the rotational band in Ref.~\cite{STS15},
and it has been successfully applied to nuclei
in the rare earth region employing the Gogny D1S interaction~\cite{STS16}.
One of the problems in the present investigation up to here is that
the calculated moment of inertia of the TSD bands in $^{163}$Lu is
considerably smaller than the measured one,
see, e.g., Fig.~\ref{fig:sLuGog0102}.
This is mainly because only one mean-field state is employed.
In fact, the calculated inertia with a single mean-field state
decreases as spin increases, for example,
${\cal J}^{(1)} \approx 51$ (64) $\hbar^2$/MeV at $I\approx 20$
while
${\cal J}^{(1)} \approx 42$ (44) $\hbar^2$/MeV at $I\approx 45$
if the cranking frequency $\omega_x=0.10$ $(0.20)$ MeV/$\hbar$ is used.
The moment of inertia increases
when the cranking frequency $\omega_x$ is increased.
Therefore, we have performed the multi-cranked configuration-mixing
calculation including four cranked HFB states with cranking frequencies,
$\omega_x=0.10,\,0.25,\,0.40,\,0.55$ MeV/$\hbar$,
in Eq.~(\ref{eq:wfProjm}) with $N_{\rm mf}=4$.

\begin{figure}[!htb]
\begin{center}
\includegraphics[width=75mm]{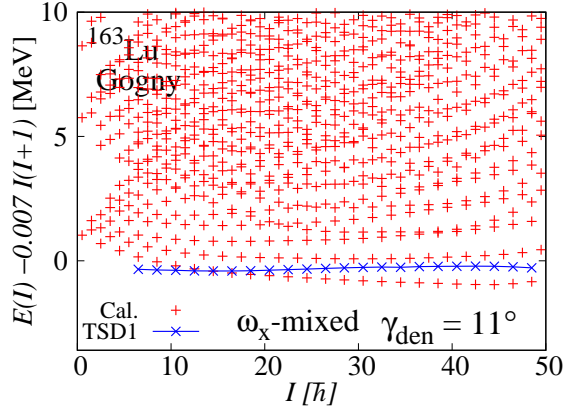}
\vspace*{-4mm}
\caption{(Color online)
Wobbling spectrum for $^{163}$Lu calculated by
the angular-momentum-projection method
with multi-cranked configuration-mixing employing
the Gogny D1S effective interaction.
The four cranked HFB states with
$\omega_x=0.10,\,0.25,\,0.40,\,0.55$ MeV/$\hbar$ are configuration-mixed.
The energy of the experimental TSD1~\cite{Jens02} is also included.
}
\label{fig:sLuGogmix}
\end{center}
\end{figure}

\begin{figure*}[!htb]
\begin{center}
\includegraphics[width=155mm]{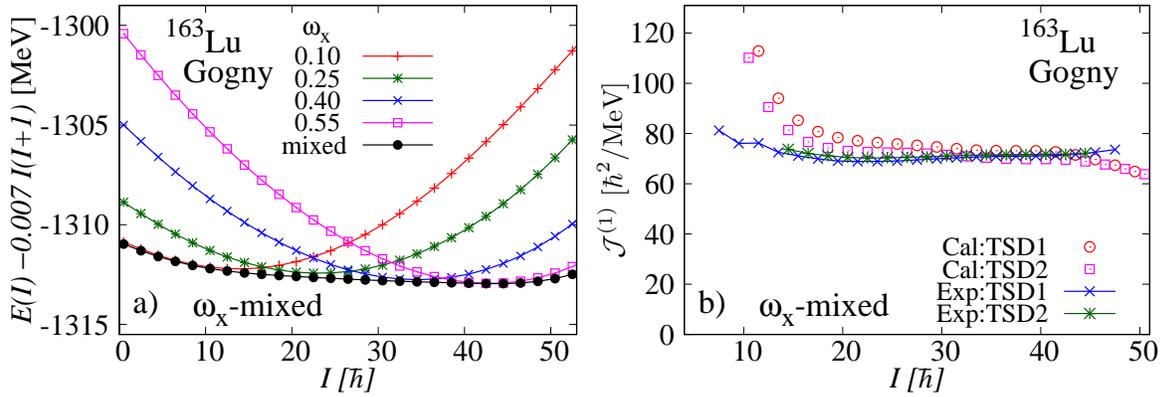}
\vspace*{-4mm}
\caption{(Color online)
Left panel: Absolute energy of the yrast band obtained
by the multi-cranked configuration-mixing
in comparison with those calculated with a single HFB state
with $\omega_x=0.10,\,0.25,\,0.40,\,0.55$ MeV/$\hbar$.
Right panel: Moments of inertia defined in Eq.~(\ref{eq:J1})
for the lowest two TSD bands
calculated by the angular-momentum-projection method
with multi-cranked configuration-mixing
corresponding to Fig.~\ref{fig:sLuGogmix}.
The experimental data are also included for comparison.
}
\label{fig:bmLuGogmix}
\end{center}
\end{figure*}

\begin{figure*}[!htb]
\begin{center}
\includegraphics[width=155mm]{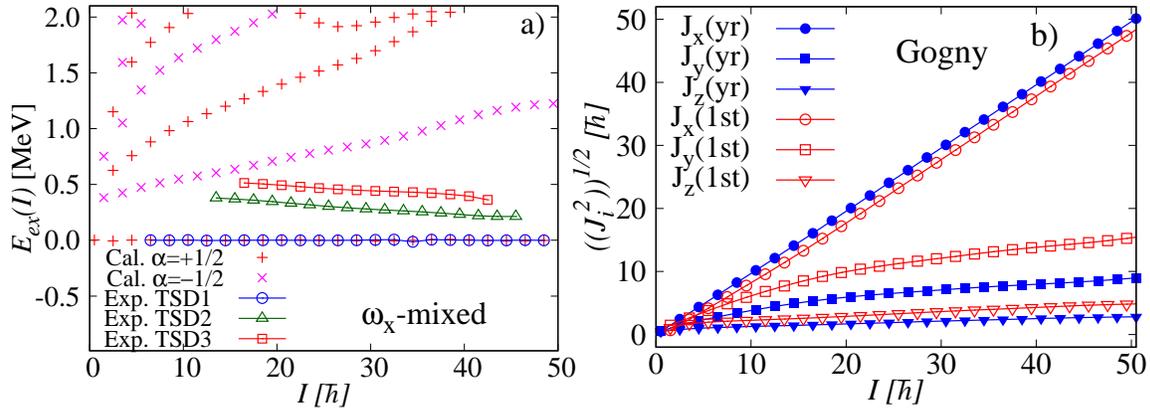}
\vspace*{-4mm}
\caption{(Color online)
The excitation energy (left) and
the expectation values of the angular-momentum vector (right) in $^{163}$Lu
calculated by the multi-cranked configuration-mixing
corresponding to Fig.~\ref{fig:sLuGogmix}.
The experimental excitation energies of TSD1, TSD2, and TSD3~\cite{Jens02}
are also included in the left panel.
}
\label{fig:ejLuGogmix}
\end{center}
\end{figure*}

The resultant spectrum is displayed in Fig.~\ref{fig:sLuGogmix}.
Clearly the lowest band nicely corresponds to the measured TSD1 band
and the result is much better than those in Fig.~\ref{fig:sLuGog0102}.
The level density of higher excited bands is considerably increased.
Even with a single mean-field wobbling bands appear
at higher excitation energy if the cranking procedure is applied,
as is shown in the previous sections (see e.g. Fig.~\ref{fig:sLunoinfcr}).
If the four mean-fields with different cranking frequencies are mixed,
various higher quasiparticle configurations are effectively included
and more excited multiple-wobbling bands are expected to come about.
This is the reason of increasing level density.
In order to see the effect of configuration-mixing we show
the calculated yrast band before and after the mixing
in the left panel of Fig.~\ref{fig:bmLuGogmix}.
It is clearly seen that the main configuration of the mixed band
smoothly changes as a function of spin.
The calculated moments of inertia ${\cal J}^{(1)}$ are compared with
the experimental data for the lowest two TSD bands
in the right panel of Fig.~\ref{fig:bmLuGogmix}.
Rather good agreement is achieved,
although the calculated moment of inertia is slightly overestimated
in the low-spin region.
Thus, the property of the rotational bands are better described
by the present multi-cranked configuration-mixing.

However, the spin-dependence of the wobbling excitation energies shown
in Fig.~\ref{fig:ejLuGogmix} are changed to increase monotonically
compared with those in Fig.~\ref{fig:ejLuGog0102},
and looks more like that of the original wobbling bands
without the alignment or of the longitudinal wobbling,
which are quite different from the experimental data.
This change can be understood; it was already shown that
if the higher cranking frequency $\omega_x$ is employed,
the resultant wobbling excitation energy increases.
Because the main component of the mixed band at higher spin states
is composed of the projected states from
the HFB state with higher cranking frequency,
it is natural that the wobbling excitation energy increases
when the spin increases.
In fact, the main component of the angular-momentum vector
is always along the $x$ axis
as is shown in the right panel of Fig.~\ref{fig:ejLuGogmix},
and the direction of the vector never change.
Namely, $ (\!( J^2_x )\!)^{1/2} - (\!( J^2_y )\!)^{1/2}$ increases
monotonically and never decreases in the calculated spin range:
This is the same behavior as in the case of the original wobbling or
of the longitudinal wobbling, although the main component is along
the short ($x$) axis in Fig.~\ref{fig:ejLuGogmix}
in contrast to the case, e.g., of Fig.~\ref{fig:ejLu30gxpnoinfcr},
where the main component is along the medium ($y$) axis.

\begin{figure*}[!htb]
\begin{center}
\includegraphics[width=155mm]{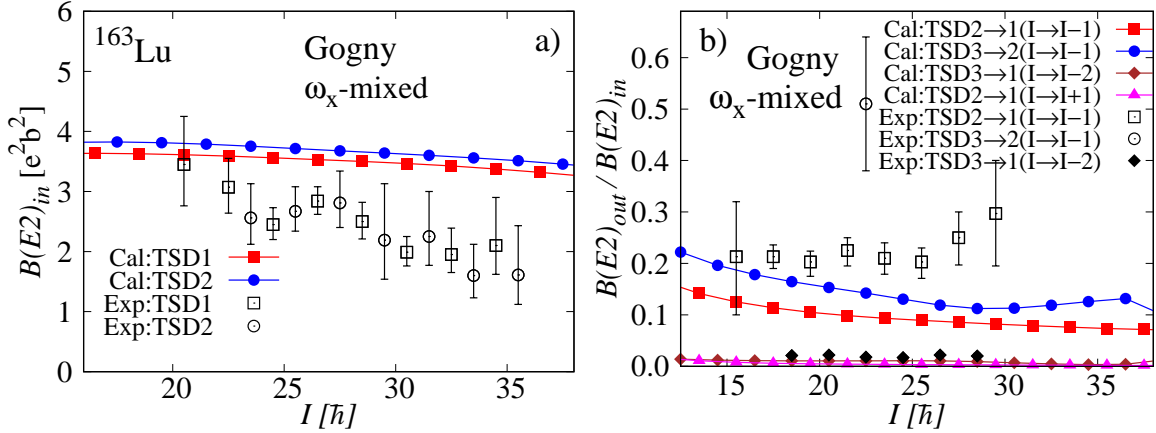}
\vspace*{-4mm}
\caption{(Color online)
The $I\rightarrow I-2$ in-band $E2$ transition probabilities (left panel)
and the $I\rightarrow I\pm 1$ out-of-band to in-band $B(E2)$ ratio (right panel),
which are calculated by the multi-cranked configuration-mixing
with the Gogny D1S interaction in $^{163}$Lu,
corresponding to Fig.~\ref{fig:sLuGogmix}.
}
\label{fig:trLuGogmix}
\end{center}
\end{figure*}

The in-band $B(E2)$ values and the out-of-band versus in-band $B(E2)$ ratios
calculated with the multi-cranked configuration-mixing
are depicted in Fig.~\ref{fig:trLuGogmix}.
The out-of-band $B(M1)$ ratio is slightly smaller at higher spins
but still about one order of magnitude larger (not shown).
Compared with the result obtained by the projection from a single HFB state
(e.g., Fig.~\ref{fig:trLuGog02std}),
the in-band $B(E2)$ values very gradually decrease with spin
because the $\beta_2$ decreases and $\gamma$ increases slightly
as functions of the cranking frequency (see Table~\ref{tab:wobgmf}).
This trend continues at higher frequency, $\omega_x \ge 0.40$ MeV/$\hbar$,
e.g. $\beta_2({\rm den})=0.423$ and $\gamma({\rm den})=13.1^\circ$,
and $\langle J_x \rangle=41.2\,\hbar$ at $\omega_x=0.55$ MeV/$\hbar$.
As for the $B(E2)_{\rm out}/B(E2)_{\rm in}$ ratios,
the results are not very different
from those obtained by the projection from a single HFB state
(e.g., in Fig.~\ref{fig:trLuGog02std}).
The spin-dependence is slightly changed;
for example, the $B(E2)_{\rm out}/B(E2)_{\rm in}$ ratios are
a little bit larger at higher spins, $I \gtsim 25$.
In any case, as it is already discussed,
the calculated selfconsistent triaxial deformation is too small
to account for the experimentally measured
$B(E2)_{\rm out}/B(E2)_{\rm in}$ ratios at high-spin states.

In this way we are not able to reproduce the observed excitation energy,
which shows the characteristic feature of the transverse wobbling
decreasing as a function of spin.
We have only tried the multi-cranked configuration-mixing
with respect to the cranking about the $x$ axis.
However, there is the possibility to mix configurations with cranking
about all three axes, although the problem of the density-dependent term
seems to appear for the Gogny (or Skyrme) interaction.
Such a possibility is interesting to be explored
with different effective interactions,
although it requires much heavier numerical efforts.

\section{Summary}
\label{sec:summary}

In the present work we have investigated the nuclear rotational motion
in triaxially deformed nuclei by employing the fully microscopic
framework of angular-momentum projection from the mean-field wave function.
As the first part of investigation,
we have concentrated on the nuclear wobbling motion,
which is quantized rotational motion of the rigid rotor.
By employing the triaxially deformed mean-field wave function,
we have confirmed that the characteristic energy spectrum
of the multiple-band structure naturally emerges
in the nucleus $^{163}$Lu by our projection calculation.
Using the deformation parameters consistent with the Nilsson-Strutinsky
calculation~\cite{BR04}, a reasonable excitation energy is obtained
for the one-phonon wobbling band but
the excitation energy of two-phonon band is too large.
Note that the excitation energies decrease as functions of spin,
i.e., the transverse wobbling behavior~\cite{FD14} comes out,
though the spin-dependence is slightly too strong
compared with the experimental data.
However, the obtained moments of inertia for the TSD bands
in $^{163}$Lu are generally too small for the employed Hamiltonian
of the Woods-Saxon potential and the schematic interaction.
To improve the moments of inertia,
we have performed infinitesimal cranking~\cite{TS16}
to include the time-odd components into the mean-fields.
The moments of inertia increase but not enough
and the excitation energy of the one-phonon band is too small
compared with the experimental data.
Therefore, we do not aim to fit the experimental data but rather perform
explorational calculations to understand the general property
of the nuclear wobbling motion from the microscopic view point.

By making use of the freedom to crank the mean-field around
all three intrinsic axes, we have investigated
how the transverse wobbling picture appears and what are its implications.
We have found that the dynamics of the angular-momentum vector
in the intrinsic coordinate frame of the mean-field are reflected
by the excitation energies of the wobbling phonon band.
Namely, the transverse-wobbling behavior,
i.e., the decrease of the wobbling excitation energy as a function of spin,
emerges when the nucleus starts to rotate around the alignment-axis of
the odd particle, which is the axis with the intermediate moment of inertia.
Increasing the spin, the rotational axis changes
to the axis with the largest moment of inertia.
The critical spin value for vanishing excitation energy
of the one-phonon wobbling band corresponds to the transition point
of changing the direction of the angular-momentum vector
from the axis with the intermediate moment of inertia
to the axis with the largest moment of inertia.
Although the moments of inertia are not explicitly introduced
in our microscopic framework,
the calculated cranking inertias (Fig.~\ref{fig:momi})
have the order ${\cal J}_y>{\cal J}_x>{\cal J}_z$ for the
medium, short, and long axes, respectively,
which is needed for transverse wobbling.
Note that the rotation around the axis with the intermediate inertia
is known to be unstable for rotation of the rigid-body without the alignment.
The unusual decreasing behavior of excitation energy is most probably
related to this fact.

As for the transition probabilities, the main conclusion is the same as
in previous work~\cite{SSM08,SS09,FD15}.
With fixed triaxial deformation, the in-band $B(E2)$ values are almost constant
and the out-of-band $B(E2)$ values decrease as a function of spin.
The comparison of these calculated $B(E2)$ values with the experimental data
suggests that the triaxial deformation of the charge density predicted
by the Nilsson-Strutinsky calculation~\cite{BR04},
$\gamma({\rm den})\approx 11-12^\circ$ in Eq.~(\ref{eq:sgammaden}),
is too small.
The decrease of the in-band $B(E2)$ and possible increase
of the ratio $B(E2)_{\rm out}/B(E2)_{\rm in}$ observed in the experimental data
may suggest that the triaxiality increases at higher spins~\cite{SS09}.
The problem of the too large out-of-band $B(M1)$ values compared with
the experimental data remains also
in the present angular-momentum-projection calculation.

For the wobbling motion in the $^{163}$Lu nucleus,
the angular-momentum-projection calculations
from the fully selfconsistent HFB mean-field
have been also performed by employing the Gogny D1S interaction.
We have found deformed HFB states
whose triaxiality is consistent with the one determined
by the Nilsson-Strutinsky calculation, i.e.,
$\gamma({\rm den})\approx 11-12^\circ$.
The resultant spectrum of the angular-momentum projection
is very similar to the one obtained
by the Woods-Saxon potential and the simple schematic interaction.
This strongly suggests that the collective wobbling
does not depend on the details of the effective interaction.
It should, however, be stressed that the more elaborate multi-cranked
configuration-mixing calculation around the short axis
can reproduce the moments of inertia of the TSD bands in a good approximation.
On the other hand, the wobbling-phonon excitation energies are not
reproduced in this configuration-mixing calculation.
In the present work, only the possibility of the one-dimensional cranking
is explored because of the problem of the density-dependent term in
the Gogny interaction.
The possibilities of cranking around all three axes should be explored
in future studies with different effective interactions.

It should be emphasized that the wobbling motion was predicted based on
the macroscopic rotor model, and the predicted properties are
nicely confirmed by our microscopic calculation.
Thus, the present study suggests that the macroscopic rotor model picture
is realized in a good approximation for triaxially deformed atomic nuclei.
It is, however, noted that a quantitative description of the wobbling motion
is not obtained in the present work, and further investigations
are necessary to achieve its fully microscopic understanding.


\vspace*{10mm}



\begin{thebibliography}{99}

\bibitem{BM75}
A.~Bohr and B.~R.~Mottelson,
{\it Nuclear Structure}, Vol.~II, Benjamin, New York (1975).

\bibitem{Mol06}
P.~M\"oller, R.~Bengtsson, B.~G.~Carlsson, P.~Olivius, and T.~Ichikawa
Phys.\ Rev.\ Lett.\ \textbf{97}, 162502 (2006).

\bibitem{VDS83}
M.~J.~A.~de Voigt,J.~Dudek and Z.~Szymanski,
Rev.\ Mod.\ Phys.\ \textbf{55}, 949 (1983).

\bibitem{Fra01}
S.~Frauendorf, Rev.\ Mod.\ Phys.\ \textbf{73}, 463 (2001).

\bibitem{Pan11}
S.~C.~Pancholi, "{\it Exotic Nuclear Excitations}",
Springer Tracts, in Modern Physics 242, Springer (2011).

\bibitem{Ode01}
S.~W.~\O{}deg\aa{}rd {\it et al.},
Phys.\ Rev.\ Lett.\ \textbf{86}, 5866 (2001).

\bibitem{NMM16}
T.~Nakatsukasa, K.~Matsuyanagi, M.~Matsuzaki, and Y.~R.~Shimizu,
Phys. Scripta {\bf 91}, 073008 (2016).

\bibitem{Fra17}
S.~Frauendorf,
Phys. Scripta, to be published, arXiv:1710.01210.

\bibitem{FM97}
S.~Frauendorf and J.~Meng,
Nucl.\ Phys.\ A \textbf{617}, 131 (1997).

\bibitem{RS80}
P.~Ring and P.~Schuck,
{\it The Nuclear Many-Body Problem}, Springer (1980).

\bibitem{TS12}
S.~Tagami and Y.~R.~Shimizu,
Prog. Theor. Phys. \textbf{127}, 79 (2012).

\bibitem{TSD13}
S.~Tagami, Y.~R.~Shimizu, and J.~Dudek,
Phys.\ Rev.\ C \textbf{87}, 054306 (2013).

\bibitem{TSD15}
S.~Tagami, Y.~R.~Shimizu, and J.~Dudek,
J.\ Phys.\ G {\bf 42}, 015106 (2015).

\bibitem{TS16} 
S.~Tagami and Y.~R.~Shimizu,
Phys.\ Rev.\ C {\bf 93}, 024323 (2016).

\bibitem{STS15}
M.~Shimada, S.~Tagami, and Y.~R.~Shimizu,
Prog. Theor. Exp. Phys. \textbf{2015}, 063D02 (2015).

\bibitem{STS16}
M.~Shimada, S.~Tagami, and Y.~R.~Shimizu,
Phys.\ Rev.\ C \textbf{93}, 044317 (2016).

\bibitem{HS95}
K.~Hara and Y.~Sun,
Int.\ J.\ Mod.\ Phys.\ E \textbf{4}, 637 (1995).

\bibitem{Sun16}
Y.~Sun, Phys.\ Scripta \textbf{91}, 043005 (2016).

\bibitem{SBD16}
J.~A.~Sheikh, G.~H.~Bhat, W.~A.~Dar, S.~Jehangir, and P.~A.~Ganai,
Phys.\ Scripta {\bf 91}, 063015 (2016).

\bibitem{FD14}
S.~Frauendorf and F.~D\"onau,
Phys.\ Rev.\ C\ {\bf 89}, 014322 (2014).

\bibitem{TSF14}
S.~Tagami, M.~Shimada, Y.~Fujioka, Y.~R.~Shimizu, and J.~Dudek,
Physica Scripta \textbf{89}, 054013 (2014).

\bibitem{BR85}
T.Bengtsson and I.~Ragnarsson,
Nucl.\ Phys.\ A \textbf{436}, 14 (1985).

\bibitem{Fra93}
S.~Frauendorf, Nucl.\ Phy.\ \textbf{557}, 259c (1993).

\bibitem{WyssPriv}
R. Wyss, private communication (2005).

\bibitem{SS09}
T.~Shoji and Y.~R.~Shimizu,
Prog.\ Theor.\ Phys.\ {\bf 121}, 319 (2009).

\bibitem{Mar79}
E.~R.~Marshalek,
Nucl.\ Phys.\ A \textbf{331}, 429 (1979).

\bibitem{TST10}
N.~Tajima, Y.~R.~Shimizu, and S.~Takahara,
Phys.\ Rev.\ C {\bf 82}, 034316 (2010).

\bibitem{D1S}
J.~F.~Berger, M.~Girod, and D.~Gogny,
Comput.\ Phys.\ Commun.\ \textbf{63}, 365 (1991).

\bibitem{KO81}
A.~Kerman and N.~Onishi,
Nucl.\ Phys.\ A \textbf{361}, 179 (1981).

\bibitem{PT62}
R.~E.~Peierls and D.~J.~Thouless,
Nucl.\ Phys.\ \textbf{38}, 154 (1962).

\bibitem{SAB90}
S.~{\AA}berg,
Nucl.\ Phys.\ A \textbf{520}, 35c (1990).

\bibitem{SchP95}
H.~Schnack-Petersen et~al.,
Nucl.\ Phys.\ A \textbf{594}, 175 (1995).

\bibitem{Hagm04}
G.~B.~Hagemann, Eur.\ Phys.\ J.\ A \textbf{20} 183 (2004).

\bibitem{BR04}
R.~Bengtsson and H.~Ryde,
Eur.\ Phys.\ J.\ A \textbf{22}, 355 (2004).

\bibitem{FD15}
S.~Frauendorf and F.~D\"onau,
Phys.\ Rev.\ C \textbf{92} 064306 (2015).

\bibitem{Shim16}
M.~Shimada, Ph.D. thesis,
Department of Physics, Kyushu University, Mar. 2016.

\bibitem{SSM08}
Y.~R.~Shimizu, T.~Shoji and M.~Matsuzaki,
Phys. Rev. C {\bf 77}, 024319 (2008).

\bibitem{Jens02} 
D. R. Jensen et al.,
Phys.\ Rev.\ Lett.\ \textbf{89}, 142503 (2002).

\bibitem{Gorg04} 
A. G\"orgen et al.,
Phys.\ Rev.\ C \textbf{69}, 031301 (2004).

\bibitem{SMM04}
Y.~R.~Shimizu, M.~Matsuzaki and K.~Matsuyanagi,
"Microscopic study of wobbling motion in Hf and Lu nuclei",
{\em Proceedings of The Fifth Japan-China Joint Nuclear Physics Symposium},
Fukuoka, Japan, pp. 317--326, arXiv:nucl-th/0404063 (2004).

\bibitem{MO04}
M.~Matsuzaki and S.-I.~Ohtsubo,
Phys. Rev. C {\bf 69}, 064317 (2004).

\bibitem{MJ78}
I.~N.~Mikhailov and D.~Janssen,
Phys.\ Lett.\ B \textbf{72}, 303 (1978).

\bibitem{Hama01} 
I. Hamamoto et al.,
Acta Phys.\ Pol.\ B \textbf{32}, 2545 (2001).

\bibitem{SM95}
Y.~R.~Shimizu and M.~Matsuzaki,
Nucl. Phys. {\bf A588}, 559 (1995).

\bibitem{DSN07}
J.~Dobaczewski, M.~V.~Stoitsov, W.~Nazarewicz, and P.-G.~Reinhard,
Phys.\ Rev.\ C \textbf{76}, 054315 (2007).

\bibitem{Rob10}
L.~M.~Robledo,
J.\ Phys.\ G \textbf{37}, 064020 (2010).

\end{thebibliography}
\end{document}